\documentclass[fleqn,usenatbib]{mnras}

\usepackage{newtxtext,newtxmath}
\usepackage{lscape}
\usepackage{graphicx}
\usepackage{pdflscape}
\usepackage[]{subfig}
\usepackage{makecell}
\usepackage[flushleft]{threeparttable}

\usepackage[T1]{fontenc}
\usepackage[flushleft]{threeparttable}

\DeclareRobustCommand{\VAN}[3]{#2}
\let\VANthebibliography\thebibliography
\def\thebibliography{\DeclareRobustCommand{\VAN}[3]{##3}\VANthebibliography}

\usepackage{graphicx}	
\usepackage{amsmath}	
\usepackage{float}
\usepackage{cleveref}

\newcommand{\correct}[1]{#1}

\title[ASKAP Galactic Centre Transients]{A pilot ASKAP survey for radio transients towards the Galactic Centre}

\author[Wang et al.]{
Ziteng Wang,$^{1,2,3}$\thanks{E-mail: zwan4817@uni.sydney.edu.au}
Tara Murphy,$^{1,3}$
David L. Kaplan,$^{4}$
Keith W. Bannister,$^{2}$
Emil Lenc,$^{2}$
James K. Leung,$^{1,2,3}$
\newauthor
Andrew O'Brien,$^{4}$
Sergio Pintaldi,$^{6}$
Joshua Pritchard,$^{1,2,3}$
Adam J. Stewart,$^{1}$
Andrew Zic,$^{5,2}$
\\
$^1$Sydney Institute for Astronomy, School of Physics, University of Sydney, Sydney, New South Wales 2006, Australia.\\
$^2$CSIRO,  Space and Astronomy, PO Box 76, Epping, New South Wales 1710, Australia\\
$^3$ARC Centre of Excellence for Gravitational Wave Discovery (OzGrav), Hawthorn, Victoria, Australia\\
$^4$Center for Gravitation, Cosmology, and Astrophysics, Department of Physics, University of Wisconsin-Milwaukee, P.O. Box 413, Milwaukee, WI 53201, USA\\
$^5$Department of Physics and Astronomy, and Research Centre in Astronomy, Astrophysics and Astrophotonics, Macquarie University, NSW 2109, Australia\\
$^6$Sydney Informatics Hub, The University of Sydney, NSW 2008, Australia
}

\date{Accepted XXX. Received YYY; in original form ZZZ}

\pubyear{2020}

\begin{document}
\label{firstpage}
\pagerange{\pageref{firstpage}--\pageref{lastpage}}
\maketitle

\begin{abstract}
We present the results of a radio transient and polarisation survey towards the Galactic Centre, conducted as part of the Australian Square Kilometre Array Pathfinder Variables and Slow Transients  pilot survey.  The survey region consisted of five fields covering $\sim265\,{\rm deg}^2$ ($350^\circ\lesssim l\lesssim10^\circ$, $\vert b\vert \lesssim 10^\circ$). Each field was observed for 12\,minutes, with between 7 and 9 repeats on cadences of between one day and four months. 
We detected eight highly variable sources and seven highly circularly-polarised sources (14 unique sources in total). Seven of these sources are known pulsars including the rotating radio transient PSR~J1739--2521 and the eclipsing pulsar PSR~J1723--2837. One of them is a low mass X-ray binary, 4U 1758--25. Three of them are coincident with optical or infrared sources and are likely to be stars. The remaining three may be related to the class of Galactic Centre Radio Transients (including a highly likely one, VAST~J173608.2--321634, that has been reported previously), although this class is not yet understood.
In the coming years, we expect to detect $\sim$40 bursts from this kind of source with the proposed four-year VAST survey if the distribution of the source is isotropic over the Galactic fields.
\end{abstract}

\begin{keywords}
radio continuum: transients -- Galaxy: centre -- radio continuum: stars
\end{keywords}



\section{Introduction}

Many types of sources exhibit variability at radio wavelengths, including pulsars, supernovae, flaring stars and X-ray binaries. The variable radio emission from these sources is often associated with high-energy astrophysical phenomenon. Studying the properties of variable sources can help us understand the physical mechanisms that cause them, as well as the properties of their local environments \citep{2015aska.confE..51F}. 
Radio transients surveys also allow us to discover new types of objects. For example, \citet{2005Natur.434...50H} discovered Galactic Centre Radio Transients (GCRTs), an unknown class of transient radio sources, in a search for transient and variable sources with the Very Large Array (VLA).

The Galactic Centre (GC) is a promising region for finding transient or variable radio sources \citep[e.g.,][]{2003ANS...324...79H}.
The GC is inhabited by stars, white dwarfs, neutron stars and stellar mass black holes, or even those objects in binary systems \citep[e.g.,][]{2003ANS...324Q..65K}.
Some of them are known to be variable or transient at radio wavelengths, such as flaring stars, pulsars and low-mass X-ray binaries \citep[e.g.,][]{2015aska.confE..51F}.
X-ray observations also support the case for searching for transient radio sources towards the GC. \citet{1993A&AS...97..149S} reported that the X-ray source density peaked towards the GC and \citet{2003ANS...324...33M} found $\sim$2000 hard X-ray sources near the GC; some of them likely associated with neutron stars.
We therefore expect a concentration of radio transients towards the GC.

There have only been a few dedicated radio transient searches towards the GC, but these searches have found a number of transients. For example, \citet{2002AJ....123.1497H, 2003ANS...324...79H, 2005Natur.434...50H, 2009ApJ...696..280H} searched for radio transients at 330\,MHz toward the GC with the VLA and found three unclassified objects that they called GCRTs. \citet{2016ApJ...833...11C} searched for short timescale transients in archival VLA  data and found two transient (but possibly spurious) sources without convincing explanations. \citet{2020ApJ...905..173Z} detected 82 variable or transient Galactic centre compact radio sources and argued that  some of the sources may be undiscovered pulsars.

Some transient and variable sources are known to be circularly polarised. 
Flaring stars can give off circular polarised emission by plasma emission or electron cyclotron maser emission \citep[e.g.,][]{2017ApJ...836L..30L}.
Pulsars can emit circular polarised emission \citep[e.g.,][]{2018MNRAS.474.4629J} but the origin remains unclear \citep[e.g.,][]{1990ApJ...352..258R, 2004MNRAS.352..915M}.
\citet{2010ApJ...712L...5R} found that bursts from GCRT~J1745--3009 can reach as high as 100 per cent circular polarisation. 
In contrast, fewer than 0.1 per cent of radio sources are circularly polarised at more than a few percent of their total intensity, which makes it easier to find polarised sources \citep[e.g.,][]{2018MNRAS.478.2835L}, especially in crowded regions.
Previous polarisation searches have identified new sources, including a new millisecond pulsar \citep[PSR J1431--6328;][]{2019ApJ...884...96K}, the first brown dwarf discovered at radio wavelengths \citep{2020ApJ...903L..33V}, and previously undetected flaring stars \citep{2021MNRAS.502.5438P, 2021NatAs...5.1233C}.

\begin{figure*}
    \centering
    \includegraphics[width=0.8\textwidth]{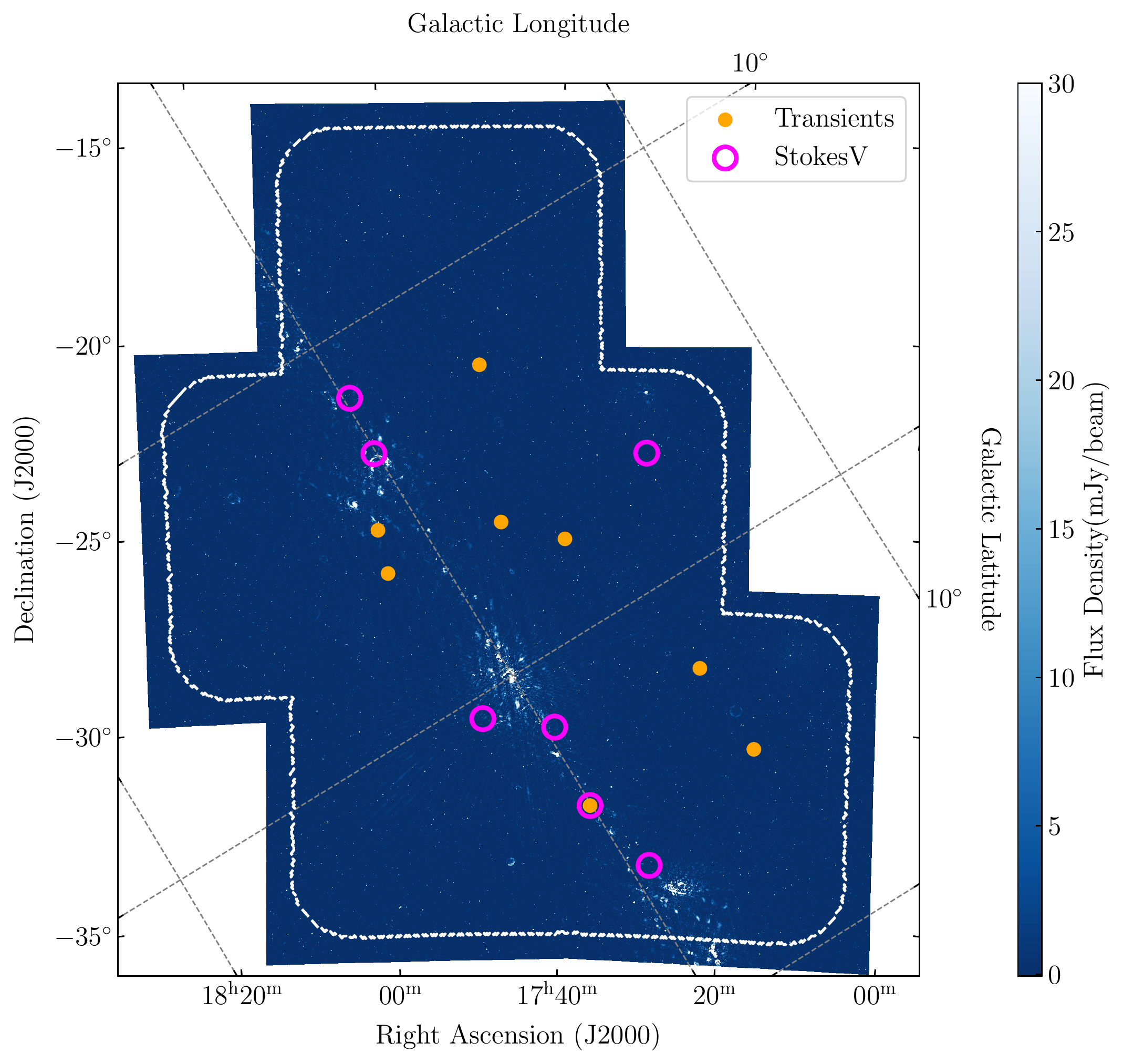}
    \caption{A mosaicked image of the RACS-low Low Galactic Latitude region. White dashed line shows the boundary of VAST Low Galactic Latitude region.
    Due to lack of neighboring fields, the VAST coverage is smaller than RACS-low though they share the same footprint.
    A grid of Galactic coordinates is shown in grey dashed lines.
    The transient and circularly polarised (Stokes V) sources we found are shown as blue circles and magenta hollow circles,  respectively.}
    \label{fig:Lowlat_mosaic}
\end{figure*}

The Australian Square Kilometre Array Pathfinder \citep[ASKAP;][]{2008ExA....22..151J, 2021PASA...38....9H} was designed as a rapid wide-field survey instrument.
It is a radio interferometer with a wide field-of-view ($\sim$30\,deg$^2$) and its sensitivity can reach a typical rms  of 0.24\,mJy in a 12\,min integration.
ASKAP's wide field-of-view and high instantaneous sensitivity improve the effectiveness for transient surveys, notably the ASKAP survey for Variables and Slow Transients \citep[VAST;][]{2013PASA...30....6M,2021PASA...38...54M}.
VAST will investigate the dynamic radio sky on timescales from seconds to months. 
During 2019--2020, VAST was allocated 100 hours observing time to conduct a Phase I of the Pilot Survey \citep[VAST-P1;][]{2021PASA...38...54M}. 
VAST-P1 consists of six survey regions covering 5131\,deg$^2$. 
Observations for VAST-P1 commenced in mid-2019 and finished in mid-2020. 
In this paper, we present the results from the Pilot Survey searching for variable and circularly polarised sources in the low Galactic latitude region (see Figure~\ref{fig:Lowlat_mosaic}). 
In $\mathsection$\ref{sec:qc}, we describe the observations and check the data quality; in $\mathsection$\ref{sec:search}, we present the methods for searching variable and circular polarised sources and we summarize the details for all the candidates; and in $\mathsection$\ref{sec:res}, we discuss the candidates from our searches.

\section{Observations and Quality Checks}\label{sec:qc}

In this paper, we used the data from two ASKAP surveys: the Rapid ASKAP Continuum Survey low \citep[RACS-low;][]{2020PASA...37...48M} and VAST-P1. Both surveys were conducted at a central frequency of 887.5\,MHz with a bandwidth of 288\,MHz. All four instrumental polarisation products (XX, XY, YX, and YY) were recorded for both surveys to allow images to be made in four Stokes parameters (I, Q, U, and V).
We used \textit{combined} images in our search to improve the sensitivity in the overlapped regions. \textit{Combined} images are made by mosaicking the individual field images together per epoch (see \citealt{2021PASA...38...54M} for details).
The VAST-P1 survey incorporates RACS-low as its first epoch, and we therefore refer to RACS-low as ``epoch 00'' in this paper.
\correct{We used the {\sc Selavy} source finding software \citep{2012PASA...29..371W} with its default settings to produce source catalogues for the images. {\sc Selavy} models source components with 2D Gaussians. The uncertainty for flux density measurements is based on the uncertainty in the Gaussian fitting and the local noise.}

\subsection{RACS-low}\label{subsec:RACS}

There are five tiles near the Galactic Centre in RACS-low: 1724--31A, 1739--25A, 1752--31A, 1753--18A and 1806--25A. The median RMS noise of these five tiles was $\sim 0.36\,{\rm mJy}/{\rm beam}$, which is higher than the typical noise of RACS-low due to the bright and diffuse emission near the Galactic Centre.
The observations were conducted between 2019 April 25 and 2019 April 28.

We used the early processing RACS-low data in this paper. Further improvements have been applied to the published RACS-low images and source catalogues. A detailed discussion is in \citet{2020PASA...37...48M} and \citet{2021PASA...38...58H}.

\begin{figure}
    \centering
    \includegraphics[width=\columnwidth]{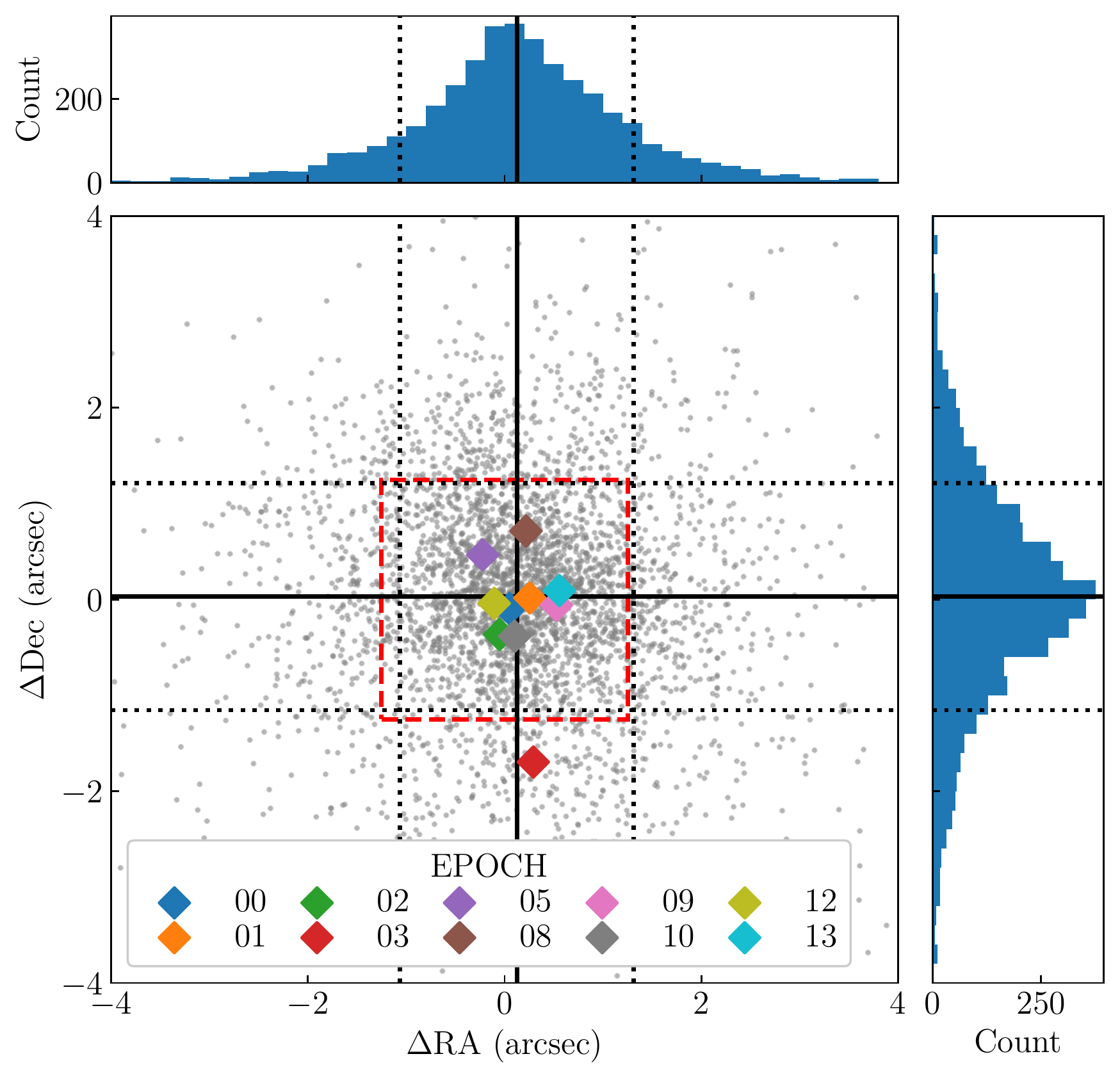}
    \caption{Astrometric accuracy for bright compact isolated sources in region 5 of pre-release RACS-low (epoch 00) and VAST-P1 compared to sources from the published RACS catalogue. Red dashed box shows the image pixel size ($2.5\times2.5$\,arcsec). 
    The offsets for each matched pair are shown in gray circles.
    We show the median offset for each epoch as coloured diamonds.}
    \label{fig:qc_astrometry}
\end{figure}
\begin{figure}
    \centering
    \includegraphics[width=\columnwidth]{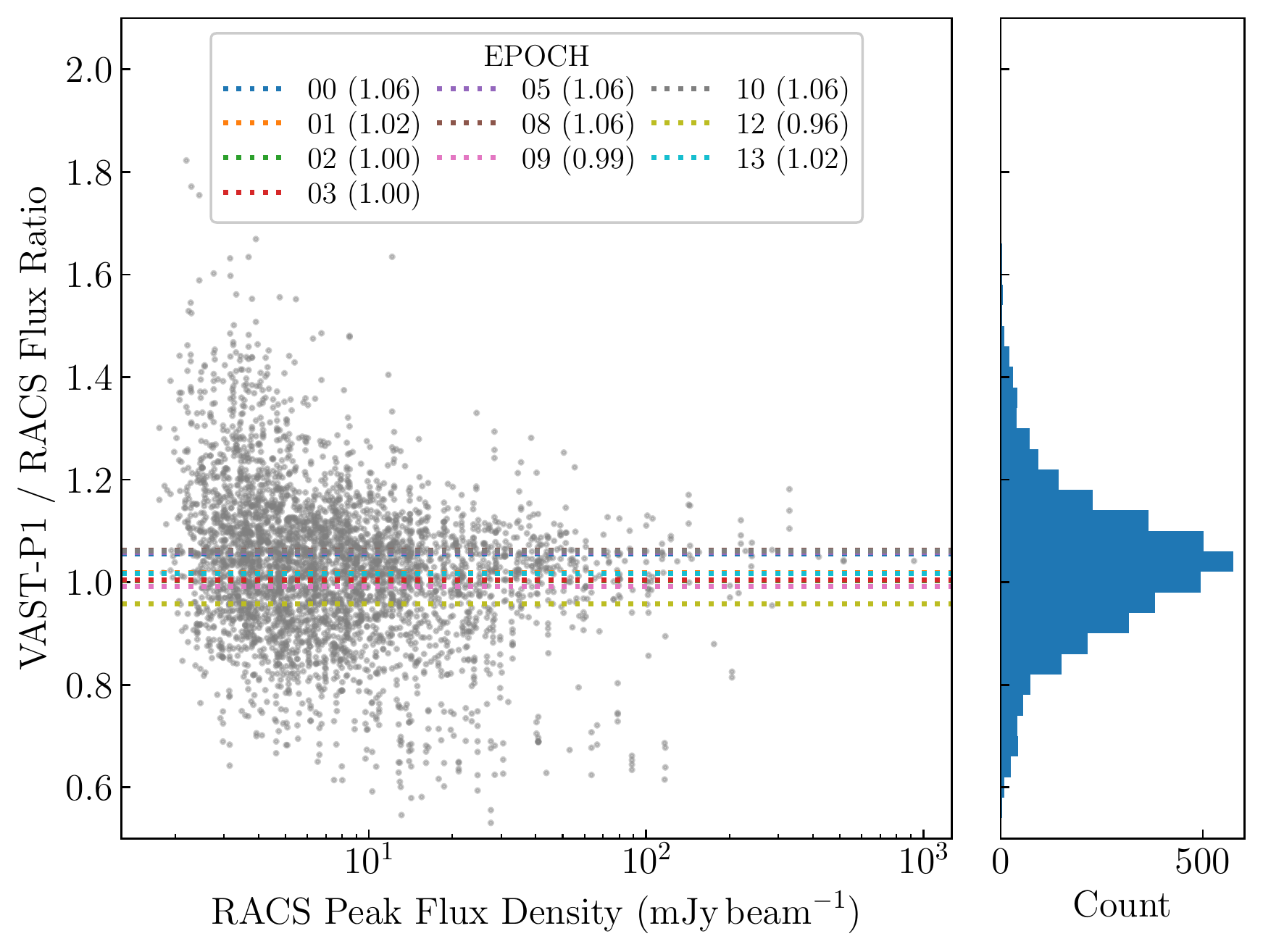}
    \caption{Flux density scale comparison for sources in region 5, comparing those measured in pre-release RACS-low (epoch 00) and VAST-P1 to RACS. We show the median ratio for each epoch as dotted lines. The ratios for each matched pair are shown in gray circles. The overall median ratio is 1.03 with a scatter of 0.15.}
    \label{fig:qc_fluxscale}
\end{figure}

We performed quality checks of the astrometric accuracy and flux density scale for the pre-release RACS-low data. We extracted bright (SNR $\geq 7$), compact\footnote{Compactness follows the definition by \citet{2021PASA...38...58H} of an integrated to peak flux ratio of $S_I/S_P < 1.024 + 0.69\times{\rm SNR}^{-0.62}$.} sources with {\sc Selavy} that are isolated by a minimum of 150\,arcsec from other sources. 
\correct{These sources have low positional errors ($\lesssim$ 1.5\,arcsec), are less likely to be spurious detections or artefacts, and are free from contamination from close neighbours, which enables us to get a robust astrometry and flux density scale comparison.}
We then crossmatched them with bright compact isolated sources (1145 sources within low Galactic latitude region) in the published RACS catalogue \citep{2021PASA...38...58H}. 
There were 482 sources matched using a 10 arcsec crossmatch radius.
The median and standard deviation of the positional offsets are $0.01\pm1.02$\,arcsec in right ascension and $-0.12\pm0.90$\,arcsec in declination (see Figure~\ref{fig:qc_astrometry}). 
The flux density ratio is 1.05 with a standard deviation of 0.12 (see Figure~\ref{fig:qc_fluxscale}).

\subsection{VAST-P1}

VAST-P1 uses the same tiling footprint, frequency, and bandwidth as RACS-low. 
There are 13 epochs in total for VAST-P1, but seven of them only have partial coverage of the full survey footprint (see \citealt{2021PASA...38...54M} for details).
There are five VAST-P1 tiles covering $\sim$265\,deg$^2$ near the Galactic Centre region. 
Each tile was observed with an integration time of $\sim12\,{\rm min}$. The median RMS noise of the five tiles we used is $\sim0.39\,{\rm mJy}/{\rm beam}$. 
All observations were processed using standard procedures in the {\sc ASKAPsoft} package \citep{ASKAPSOFT.2016,2019ascl.soft12003G} and {\sc Selavy} \citep{2012PASA...29..371W} was used for source finding and flux measurement.

We performed the quality checks on VAST-P1 data for the Galactic Centre region in the same way as for the pre-release RACS-low data. 
The observation for field 1724--31A in epoch 09 had poor calibration, which resulted in astrometry errors and reduced flux densities, and so we excluded this observation in the rest of our analysis. 
All nine epochs have a median VAST-P1/RACS-low flux density ratio between 0.96 and 1.06 (within 10 per cent, see Figure~\ref{fig:qc_fluxscale}). The overall ratio is 1.03 with a 1$\sigma$ scatter of 0.15. 
In our analysis, we selected variability thresholds such that the overall flux variation between epochs is not significant (see Section~\ref{subsec:variable_search}).
The astrometric accuracy for VAST-P1 is shown in Figure~\ref{fig:qc_astrometry}. The median and standard deviation of the positional offsets are $0.15\pm1.20$ arcsec in right ascension and $0.04\pm1.22$ arcsec in declination. We note that the offset for epoch 03 was larger compared to other epochs, but still within a single image pixel  ($<$2.5\,arcsec).

\section{Search Methodology}\label{sec:search}

\subsection{Variable Search}\label{subsec:variable_search}

We performed a search for variable sources using the VAST pipeline \citep{2021PASA...38...54M, 2021arXiv210105898P}. 
For a given source, in an epoch in which it was not detected, we used the synthesized beam of that epoch to perform forced fitting at the source position to get a \textit{forced} measurement.
We found 46\,732 unique sources in all fields and epochs. We excluded 6\,776 sources with only one measurement (most of such sources appear in the RACS-low footprint, but not the VAST footprint, see Figure~\ref{fig:Lowlat_mosaic}) and 12\,637 sources that were only detected in one epoch (but with $>1$ measurement) but had an SNR $<$ 7.0 to avoid false detections. This resulted in 29\,410 unique sources to analyse for variability.
We used the modulation index ($V$) to measure the degree of variability, and the reduced chi-square ($\eta$) statistic to show the significance of the measured variability  \citep{2019A&C....27..111R}. They are defined as
\begin{align}
V &= \cfrac{\sigma_S}{\bar{S}} \\
    \eta &= \cfrac{1}{N-1}\sum_{i=1}^N \cfrac{\left(S_i - \bar{S}_{\rm wt}\right)^2}{\sigma_i^2}
\end{align}
where $S_i$ is the flux density in epoch $i$, $\bar{S}$ is the mean flux density, $\sigma_i$ is the detection uncertainty, and $\bar{S}_{\rm wt}$ is the weighted mean flux density which is given by
\begin{equation}
    \bar{S}_{\rm wt} = 
    \cfrac{\sum_{i=1}^n \left(S_i/\sigma_i^2\right)}
    {\sum_{i=1}^n \left(1/\sigma_i^2\right)}
\end{equation}

\begin{figure}
    \centering
    \includegraphics[width=0.95\columnwidth]{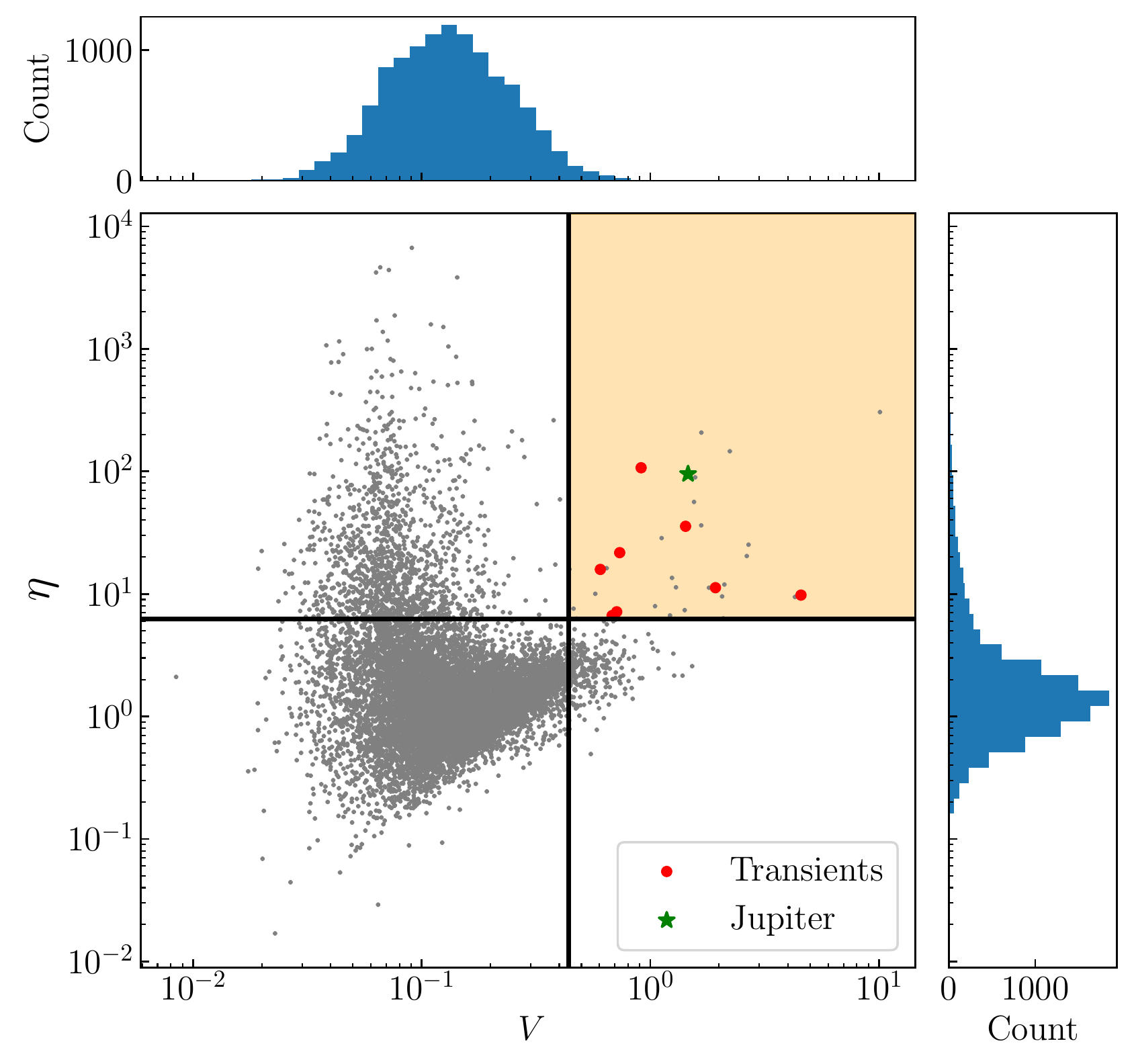}
    \caption{Transient phase-space ($\eta$-$V$) plot for all sources. Red dots show the variable sources detected in our analysis and the green star shows the Jupiter detected in the survey by chance. Black lines represent the minimum threshold of variability statistics (vertical line: $V=44.0\%$ and horizontal line: $\eta=6.23$) we used among fields. The gray dots show the sources that does not meet our selection criteria or were ruled out by manual inspection.} 
    \label{fig:eta_V_plot}
\end{figure}

Before determining the variability thresholds, we excluded extended sources and sources likely to be artefacts by removing sources that satisfied any of the following criteria:
\begin{itemize}
    \item The source was extended (the ratio of integrated flux to peak flux $>$ 1.5);
    \item The source contained multiple components;
    \item The source was close to other sources (separation $<$ 30 arcsec\correct{, which is about two times the size of the point spread function for VAST-P1});
    \item The source was close to the edges of the image ($<$ 2$^\circ$\correct{, where the typical rms is much higher, $\sim$ 1.5\,mJy, compared to the image centre}).
\end{itemize}
When calculating the thresholds, we excluded outliers (using a 3$\sigma$ threshold) and fit the $\eta$ and $V$ distributions with Gaussians in logarithmic space \citep{2019A&C....27..111R} to calculate the means ($\mu$) and standard deviations ($\sigma$). 
We used a $2\sigma$ (i.e.: $\mu + 2\sigma$) threshold to select our candidates ($V > 0.44$, $\eta > 6.23$).
\correct{The $2\sigma$ threshold enabled us to pick sources that are highly variable and also makes the number of candidates manageable for manual inspection (see \citealt{2021PASA...38...54M}).} 
This resulted in 35 sources in our candidate space, as shown in the  $\eta$-$V$ plot in Figure~\ref{fig:eta_V_plot}.

\begin{figure*}
    \centering
    \includegraphics[width=0.98\textwidth]{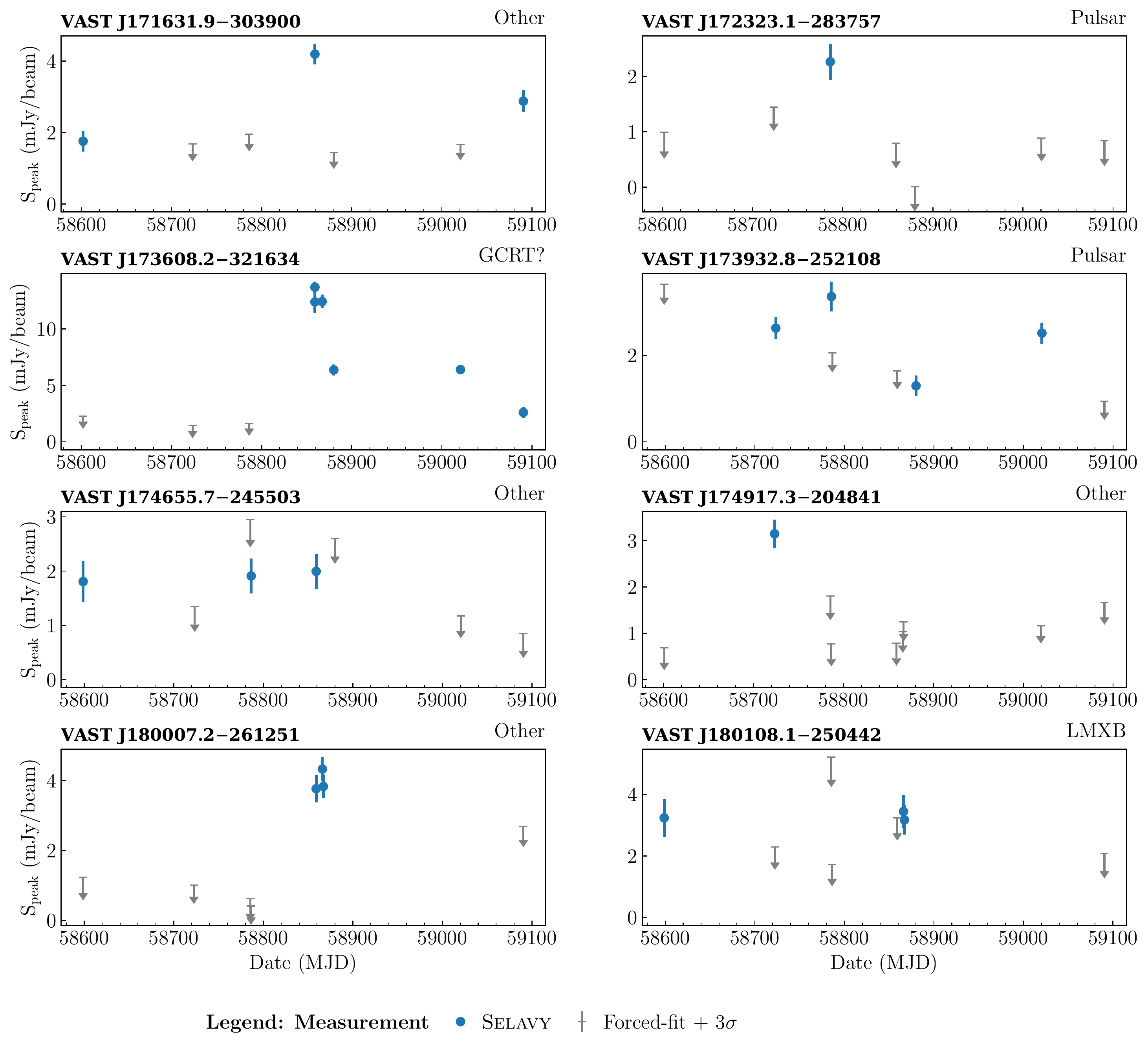}
    \caption{Lightcurves of the variables identified in VAST-P1 Low Galactic Latitude region. Blue circles are peak flux density measurements from \textsc{Selavy}. Gray lines show the $3\sigma$ upper-limits of the forced-fitted flux density for images where there was no \textsc{Selavy} detection.}
    \label{fig:lightcurve}
\end{figure*}

After manually inspecting images and lightcurves for the 35 candidates, we found eight variable sources (see Table~\ref{tab:results}). 
Jupiter moved across the field of view in one observation and was identified as a variable (green star in Figure~\ref{fig:eta_V_plot}).
The other candidates were rejected because they were coincident with either imaging artefacts or components with extended sources.
The lightcurves of the eight variable sources are presented in Figure~\ref{fig:lightcurve}. We analyse each of these sources in $\mathsection$\ref{sec:res}. The variability statistics and the properties of the sources are listed in Table~\ref{tab:results}.

\subsection{Polarisation Search}\label{subsec:cp_search}
We also performed a search for circularly polarised sources using the method described by \citet{2021MNRAS.502.5438P}. We used {\sc Selavy} with default setting for source finding and flux measurement in the Stokes V images. We extracted 1\,835 positive components from the normal images and 2\,443 negative components from the inverted images. 
We crossmatched the combined 4\,278 components against the Stokes I components using a radius of 15~arcsec, to identify the corresponding component in Stokes I. 
We searched for compact polarised sources only, so we excluded extended components (the ratio of integrated flux to peak flux $> 1.5$) and components with siblings.
We crossmatched all components in different epochs again using a 15~arcsec crossmatch radius which resulted in 1\,740 unique sources and we selected 95 sources with fractional polarisation $f_p = |V|/I$ greater than 6 per cent for manual inspection.
The threshold we chose is 10 times the median circular polarisation leakage in RACS-low ($\sim$0.6 per cent, see \citealt{2021MNRAS.502.5438P}) \correct{to minimise the possibility of selecting leakage in our sample}.

\begin{figure}
    \centering
    \includegraphics[width=0.95\columnwidth]{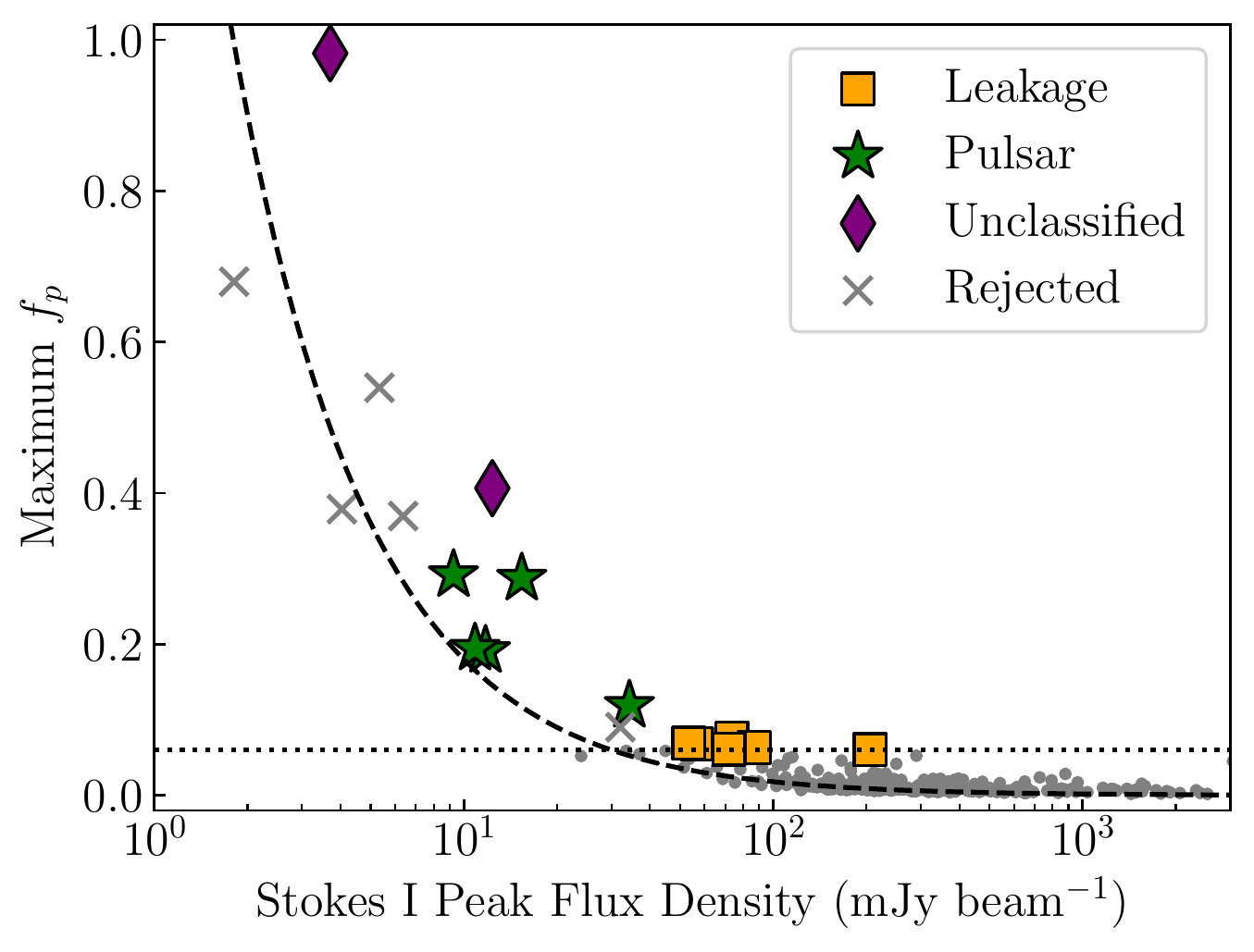}
    \caption{Maximum circular polarisation ratio of sources with respect to corresponding Stokes I flux measurements. Black dashed line shows the $5\sigma /I$, where $\sigma$ is approximate noise ($\sim0.36$\,{\rm mJy/beam}) among all fields and epochs in Stokes V images. Black dotted line represents the threshold (6 per cent) we used to pick up candidates. The known pulsars found in the search are shown in green stars. Orange squares are possible leakage from bright sources. Two unclassified sources are shown in purple diamonds. Gray crosses denote the sources rejected after the manual inspection. The details of the sources are listed in Table~\ref{tab:results}.}
    \label{fig:stokesV_plot}
\end{figure}

Figure~\ref{fig:stokesV_plot} shows all highly circular polarised sources we found.
Five of them were identified as known pulsars, six sources are associated with leakage from bright Stokes I sources (all six sources are at the edge of the \textit{combined} image, where the leakage is expected to be higher, see e.g., \citealt{2021MNRAS.502.5438P}), and there are two unclassified sources with no known pulsars or stars matched (we show the details in Table~\ref{tab:results}). All other sources (the grey crosses in Figure~\ref{fig:stokesV_plot}) were either imaging artefacts caused by leakage of the bright Stokes I sources or associated with extended Stokes I sources such as radio galaxies, planetary nebula and an \ion{H}{ii} region.

\begin{table*}
  \caption{Highly variable and circularly polarised sources identified in the VAST-P1 Low Galactic Latitude region. The coordinate of each source is given as the weighted average of all \textsc{Selavy} detections, where the weight is the inverse square of the positional error. $\sigma_\text{pos}$ is the averaged positional uncertainty. $\eta$ and $V$ are the variability parameters described in the text. nE gives the number of epochs (observations) that cover the source location. nD gives the number of detections. $|{\rm V}|/{\rm I}$ is the ratio of Stokes V to Stokes I flux density measured in the epoch for which this is a maximum, or the most constraining $3\sigma$ upper limit in the case of non-detections in Stokes V. Method shows the way the source was selected.}
  \label{tab:results}
    \begin{tabular}{lcccrrccrrcl}
\hline
Source Name & RA & Dec & $\sigma_\text{pos}$ & $\eta$ & $V$ & nE & nD & \multicolumn{1}{c}{$S_\text{max}$} & \multicolumn{1}{c}{$|{\rm V}|/{\rm I}$} & Method & ID \\
 & (J2000) & (J2000) & ($''$) &  &  &  &  & (mJy/ beam) & & & \\
\hline
\multicolumn{2}{l}{Pulsars} \\
\hline
VAST J172323.1$-$283757 & 17:23:23.1 & $-$28:37:57 & 0.6 & 9.78 & 4.56 & 7 & 1 & 2.3$\pm$0.3 & $<0.47$ & var & PSR J1723$-$2837\\
VAST J173021.7$-$230431 & 17:30:21.7 & $-$23:04:31 & 0.5 & 28.66 & 0.19 & 8 & 8 & 12.4$\pm$0.4 & 0.19 & cir & PSR J1730$-$2304\\
VAST J173932.8$-$252108 & 17:39:32.8 & $-$25:21:08 & 0.6 & 15.84 & 0.60 & 8 & 4 & 3.4$\pm$0.3 & $<0.32$ & var & PSR J1739$-$2521\\
VAST J174033.8$-$301542 & 17:40:33.8 & $-$30:15:42 & 0.5 & 2.00 & 0.10 & 9 & 9 & 11.6$\pm$1.2 & 0.29 & cir & PSR J1740$-$3015\\
VAST J174913.5$-$300235 & 17:49:13.5 & $-$30:02:35 & 0.6 & 5.42 & 0.14 & 7 & 7 & 10.9$\pm$0.9 & 0.19 & cir & PSR J1749$-$3002\\
VAST J180119.8$-$230444 & 18:01:19.8 & $-$23:04:44 & 0.5 & 2.16 & 0.08 & 8 & 8 & 43.2$\pm$1.7 & 0.12 & cir & PSR J1801$-$2304\\
VAST J180351.4$-$213706 & 18:03:51.4 & $-$21:37:06 & 0.5 & 15.10 & 0.14 & 9 & 9 & 21.4$\pm$0.6 & 0.29 & cir & PSR J1803$-$2137\\
\hline
\multicolumn{2}{l}{Low Mass X-ray Binaries} \\
\hline
VAST J180108.1$-$250442 & 18:01:08.1 & $-$25:04:42 & 0.6 & 6.65 & 0.68 & 8 & 3 & 3.4$\pm$0.5 & $<0.32$ & var & 4U 1758$-$25\\
\hline
\multicolumn{2}{l}{Galactic Centre Radio Transients} \\
\hline
VAST J173608.2$-$321634 & 17:36:08.2 & $-$32:16:34 & 0.5 & 107.00 & 0.91 & 9 & 6 & 13.7$\pm$0.5 & 0.41 & var/cir & ?\\
\hline
\multicolumn{2}{l}{Other sources} \\
\hline
VAST J171631.9$-$303900 & 17:16:31.9 & $-$30:39:00 & 0.6 & 21.70 & 0.74 & 7 & 3 & 4.2$\pm$0.3 & $<0.26$ & var & Star?\\
VAST J172841.2$-$334548 & 17:28:41.2 & $-$33:45:48 & 0.6 & 7.67 & 1.66 & 7 & 1 & 3.7$\pm$0.6 & 0.98 & cir & --\\
VAST J174655.7$-$245503 & 17:46:55.7 & $-$24:55:03 & 0.6 & 7.14 & 0.71 & 8 & 3 & 2.0$\pm$0.3 & $<0.54$ & var & Star?\\
VAST J174917.3$-$204841 & 17:49:17.3 & $-$20:48:41 & 0.5 & 11.24 & 1.93 & 9 & 1 & 3.1$\pm$0.3 & $<0.35$ & var & --\\
VAST J180007.2$-$261251 & 18:00:07.2 & $-$26:12:51 & 0.5 & 35.62 & 1.43 & 8 & 3 & 4.3$\pm$0.3 & $<0.25$ & var & Star\\
\hline
\end{tabular}

\end{table*}

\begin{figure*}
    \centering
    \includegraphics[width=\textwidth]{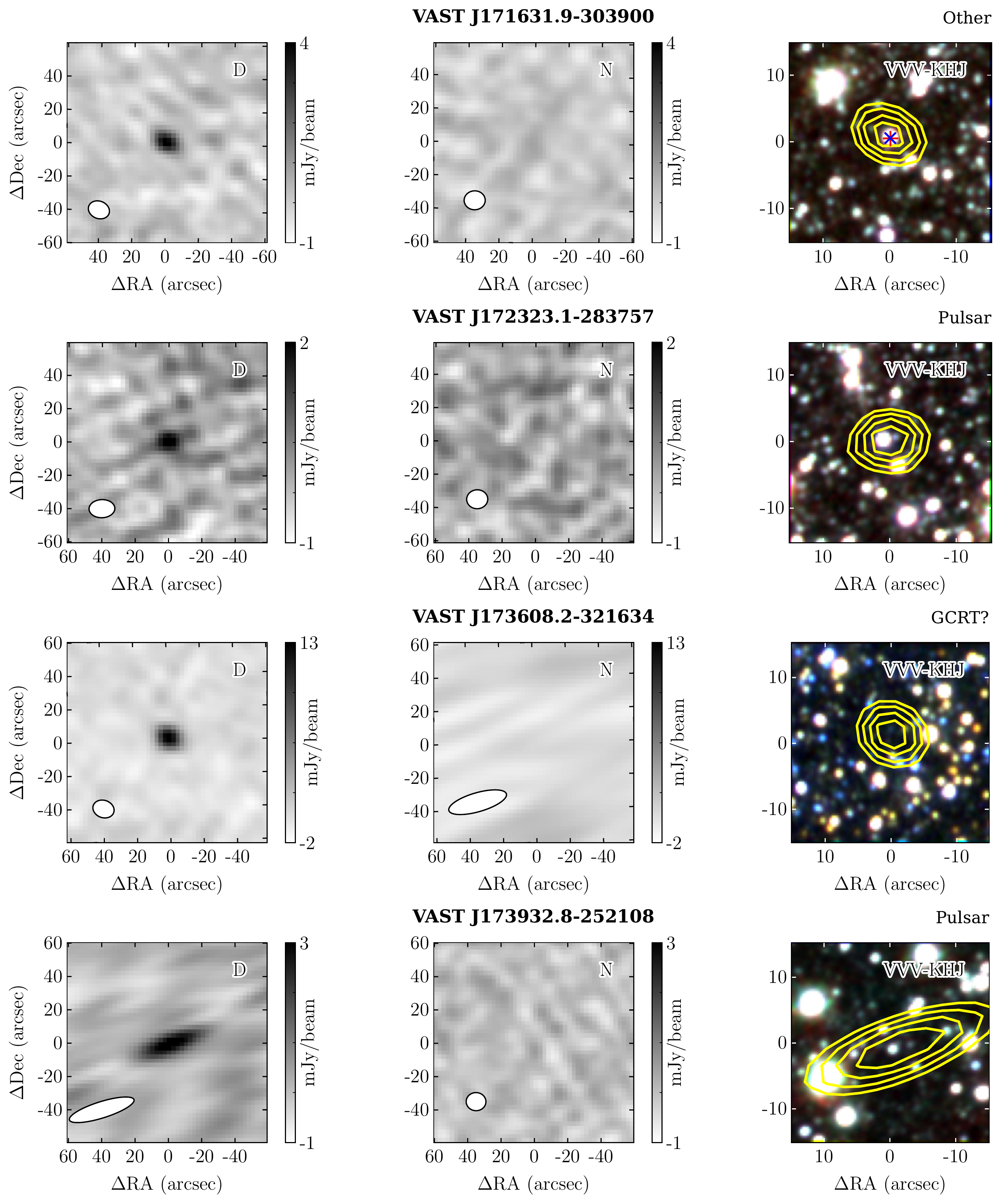}
    \caption{\correct{Images of the variable sources identified in this paper. The left panel shows the VAST Stokes I image for the epoch with the maximum flux density. The middle panel shows a non-detection epoch as a reference.} The ellipse in the lower left corner of each radio image shows the FWHM of the restoring beam. The right panels show Stokes I contours at 60, 70, 80, and 90 per cent of the peak Stokes I flux density overlaid on an RGB image of infrared data from either Vista Variables in the Via Lactea (VVV) or Two Micron All Sky Survey (2MASS) with red=\textit{K}-band, green=\textit{H}-band, blue=\textit{J}-band. 
    For sources we discussed in $\mathsection$\ref{sec:dis.other}, we also plotted their VVV and \textit{Gaia} matches in blue $\times$ and red $+$, respectively.
    All images have been centred on a frame aligned with the position of the radio source.}
    \label{fig:var_cutout}
\end{figure*}
\begin{figure*}\ContinuedFloat
    \centering
    \includegraphics[width=\textwidth]{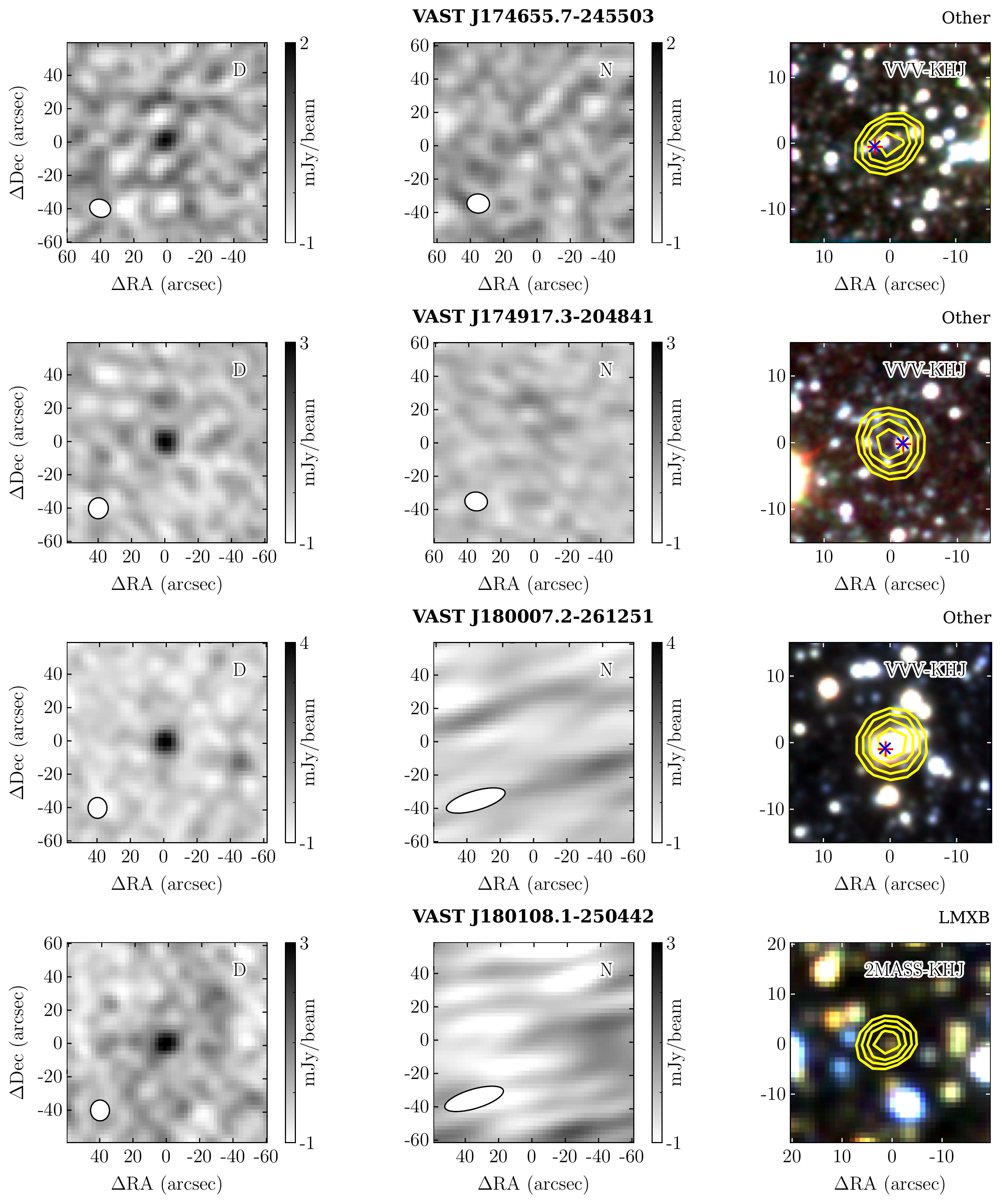}
    \caption{(continued) Images of the variable sources identified in this paper.}
\end{figure*}

\begin{figure*}
    \centering
    \includegraphics[width=\textwidth]{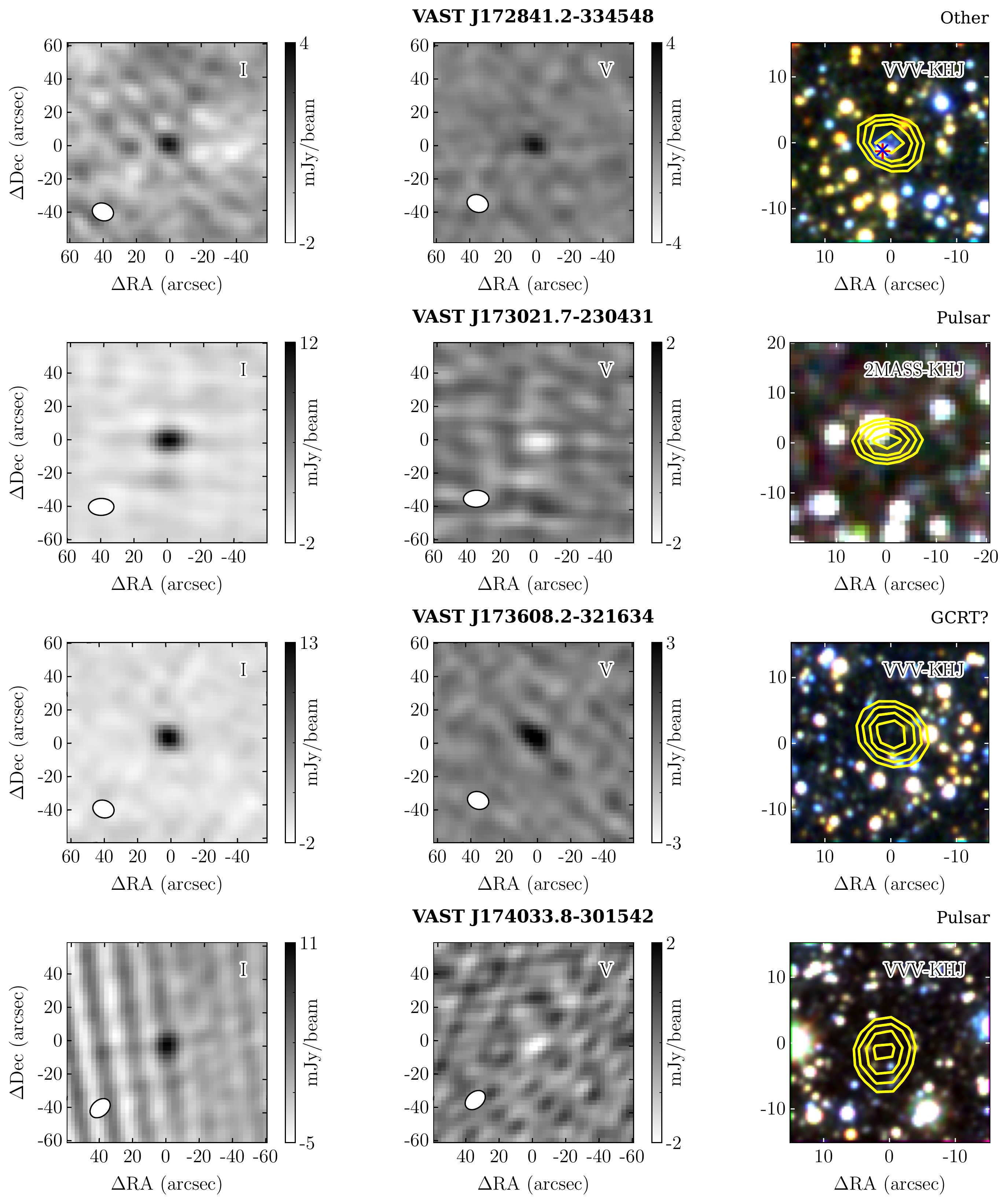}
    \caption{Images of the circularly polarised sources identified in this paper. The left panel shows the VAST Stokes I image for the epoch with the maximum flux density. The middle panel shows the Stokes V image for the same epoch where positive flux density corresponds to right handed circular polarisation and negative to left handed. The ellipse in the lower left corner of each radio image shows the FWHM of the restoring beam. \correct{The right panels show the same information as that in Figure~\ref{fig:var_cutout}.}
    All images have been centred on a frame aligned with the position of the radio source.}
    \label{fig:pol_cutout}
\end{figure*}
\begin{figure*}\ContinuedFloat
    \centering
    \includegraphics[width=\textwidth]{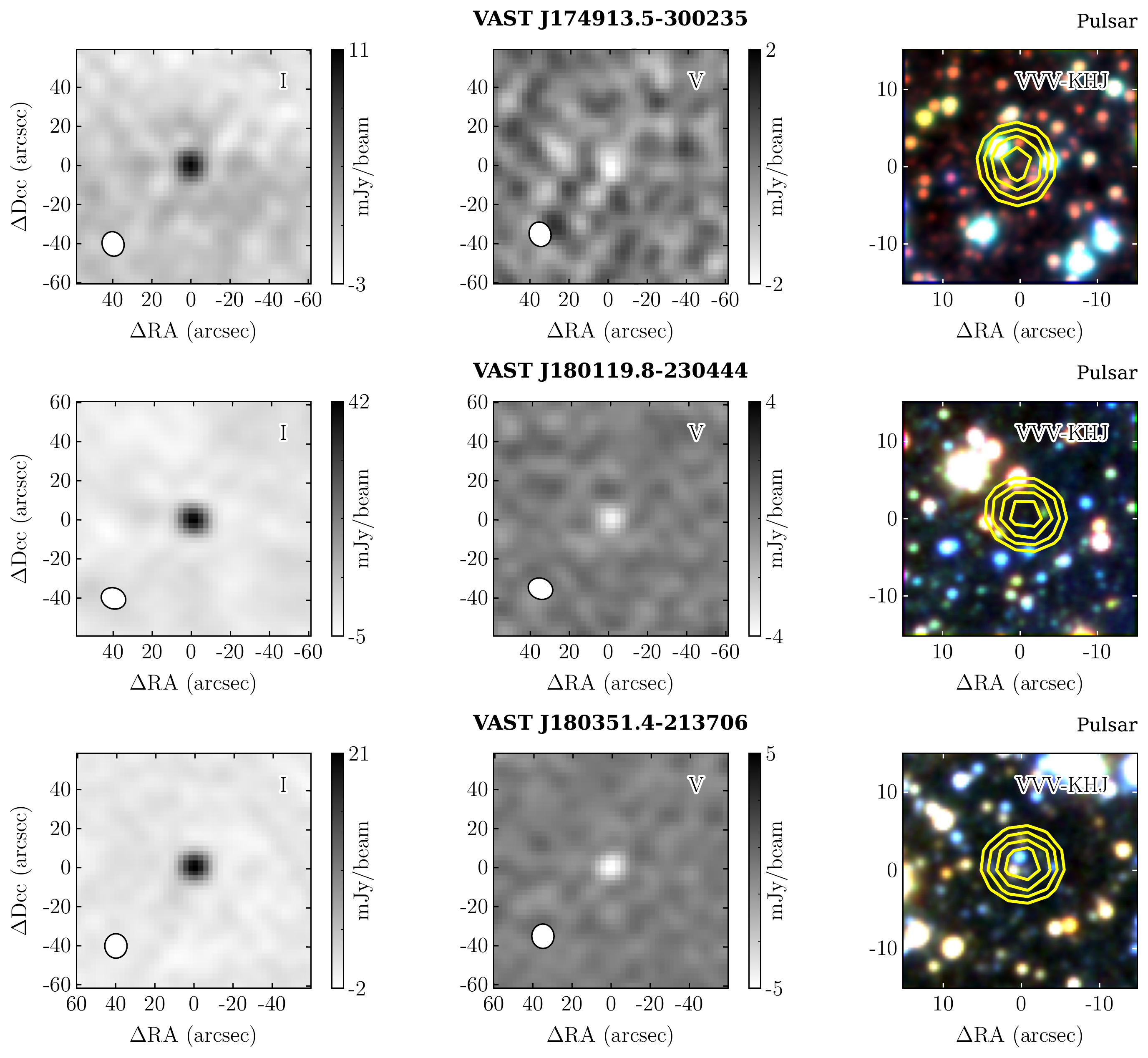}
    \caption{(continued) Images of the circularly polarised sources identified in this paper.}
\end{figure*}

\section{Results}\label{sec:res}

We found eight highly variable sources in the transient search and seven highly circular-polarised sources in the polarisation search. Together they  resulted in 14 unique sources in total.

For each source, we searched for archival detections in previous radio surveys including the  quick look images from the Karl G.\ Jansky Very Large Array Sky Survey \citep[VLASS;][]{2020PASP..132c5001L}, the TIFR GMRT Sky Survey \citep[TGSS;][]{2017A&A...598A..78I},  the GaLactic and Extragalactic All-sky MWA \citep[GLEAM;][]{2015PASA...32...25W, 2017MNRAS.464.1146H}, the NRAO VLA Sky Survey \citep[NVSS;][]{1998AJ....115.1693C} and the second epoch Molonglo Galactic Plane Survey \citep[MGPS-2;][]{2007MNRAS.382..382M}. 
We also searched for optical or infrared counterparts in \textit{Gaia} DR3 \citep{2022arXiv220605989B}, SkyMapper \citep{2018PASA...35...10W, 2019PASA...36...33O}, VISTA Variables in the Via Lactea \citep[VVV;][]{2010NewA...15..433M}, the Wide-field Infrared Survey Explorer catalogue \citep[WISE;][]{2012yCat.2311....0C} and the Two Micron All Sky Survey \citep[2MASS;][]{2006AJ....131.1163S}.
We discuss these sources individually below.

\subsection{Pulsars}
Two of our highly variable sources and five of our highly polarised sources are identified with known pulsars (\citealt{2005AJ....129.1993M}\footnote{The Australia Telescope National Facility Pulsar Catalogue: \url{https://www.atnf.csiro.au/research/pulsar/psrcat/}}, \citealt{kaplan_david_2022_6390905}\footnote{Pulsar Survey Scraper: \url{https://pulsar.cgca-hub.org/}}). 
Pulsars are one of the most highly polarised radio source classes known \citep[e.g.,][]{1988MNRAS.234..477L}.
Five pulsars (PSR J1740$-$3015, PSR J1803$-$2137, PSR J1801$-$2304, PSR J1730$-$2304, PSR J1749$-$3002) identified in our polarisation search do not appear to be remarkable, so we will not discuss them in detail.
However, two  of the pulsars we identified in the transient search are more interesting.

\begin{figure}
    \centering
    \includegraphics[width=0.9\columnwidth]{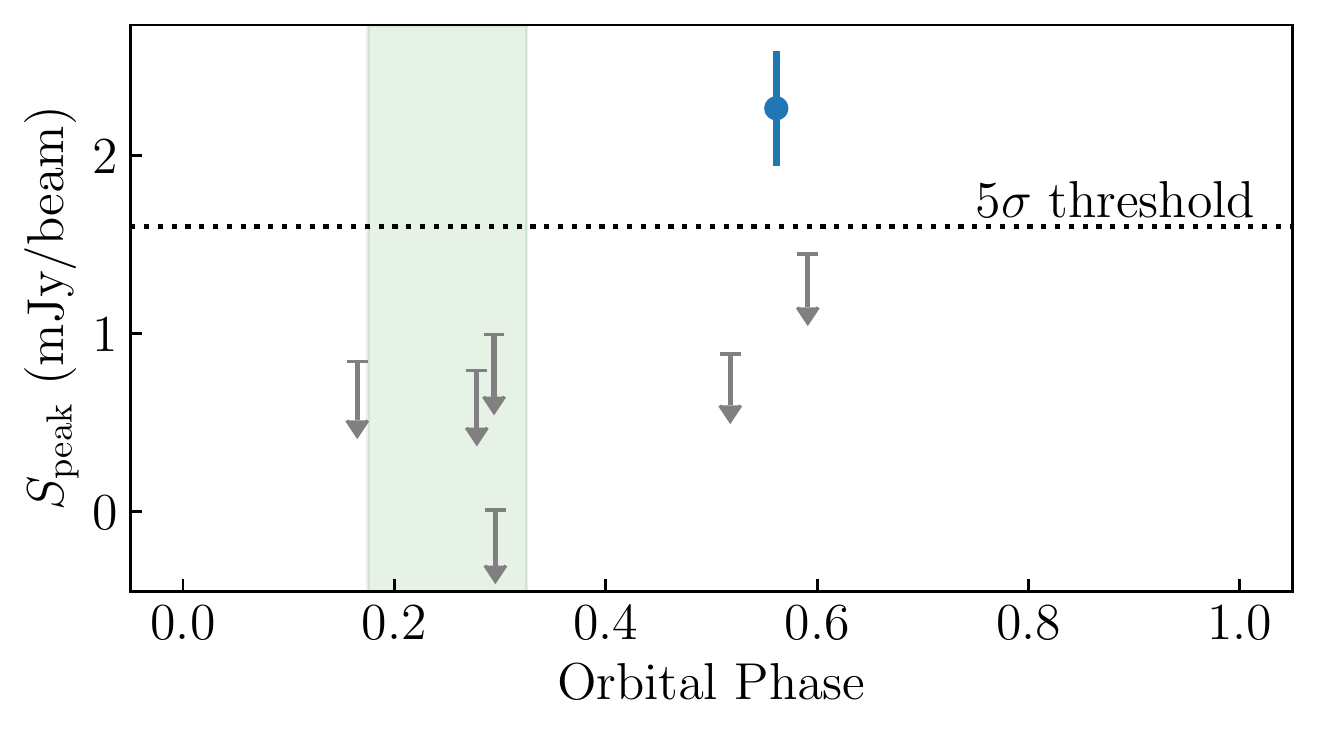}
    \caption{The VAST lightcurve for PSR~J1723$-$2837 as a function of orbital phase (the pulsar is at inferior conjunction at a phase of 0.25). All orbital phases are caculated based on the barycentred observation times. Blue circles are peak flux density measurements from {\sc Selavy} and gray lines show the $3\sigma$ upper-limits of the forced-fitted flux density. The black dotted line shows the $5\sigma$ detection threshold towards the source position. The green shaded region shows the phases when there the pulsar is eclipsed at 2\,GHz \citep{2013ApJ...776...20C}: eclipses at lower frequencies like the 900-MHz data here can have longer durations \citep{2018MNRAS.476.1968P}.}
    \label{fig:J1723_lc}
\end{figure}

\textbf{VAST J172323.1$-$283757} is identified as PSR J1723$-$2837, an eclipsing, 1.86\,ms millisecond binary radio pulsar with a $\sim$0.5\,M$_\odot$ companion \citep{2011AIPC.1357..127R, 2013ApJ...776...20C}.  Eclipsing systems with companions in this mass range are often referred to as ``redback'' pulsars \citep{2011AIPC.1357..127R}.
The orbital period of this system is $\sim$15\,hr with an eclipse duration of $\sim$15 per cent at 2000~MHz \citep[about 2.25\,hr,][]{2013ApJ...776...20C}. 
We detected this source only at one of seven epochs, at a SNR of 8.
We calculated the orbital phase based on the timing parameters  reported in \citet{2013ApJ...776...20C} and show the VAST lightcurve as a function of orbital phase in Figure~\ref{fig:J1723_lc}. The timing parameters  only predicts the orbital phase to an accuracy of $~30\,$mins ($\sim 3$ per cent of the orbit) at the time of our VAST-P1 observations. We also note that redbacks can have unpredictable orbital period variations \citep[e.g.,][]{2016ApJ...816...74B}, so we can not be sure what the precise phase is. 
Assuming the model holds, we find that only four out of six non-detections can be explained by observations during eclipse. The eclipse duration could be longer at 888\,MHz compared to 2\,GHz, but even so we have a detection at an earlier phase than a non-detection, so likely there is some other reason for the variability.  The remaining non-detections may arise from propagation effects such as interstellar scintillation, causing large variability  \citep[e.g.,][]{1990ARA&A..28..561R}.
The scintillation strength $u \approx 16$ suggests a strong scattering regime (diffractive and/or refractive scintillation). Assuming a Kolmogorov spectrum \citep[e.g.,][]{1977ARA&A..15..479R}, we calculated a diffractive scintillation bandwidth of $\Delta f_{\rm DISS} \sim 3\,$MHz and a scintillation timescale of $\sim 15\,$mins at 888\,MHz.
Though the $\Delta f_{\rm DISS}$ is much small than the observing bandwidth $\sim 288\,$MHz, our integration time is comparable to the scintillation timescale, so we may only be sampling a small number of ``scintles" and could expect significant variability. 
Assuming the spectral index ($\sim -5$) and the duty cycle ($\sim 10$ per cent) derived from \citet{2013ApJ...776...20C} remained the same, the expected flux density at VAST frequency (888\,MHz) would be $\sim 1\,$mJy. The only detection is likely to arise from scintillation, which boosts the flux density above our detection threshold.

\textbf{VAST J173932.8$-$252108} is identified as PSR J1739$-$2521, a Rotating Radio Transient (RRAT) with a $\sim$1.82\,s period \citep{2017ApJ...840....5C}.
RRATs are pulsars that show sporadic  radio bursts with separations ranging from minutes to hours \citep{2006Natur.439..817M}. 
\citet{2017ApJ...840....5C} reported a  burst rate of $\sim$23\,hr$^{-1}$ and a mean flux density for single pulses of 49\,mJy at 820\,MHz. 
The burst fluence was measured to be $\sim 3\,$Jy\,ms on average, and could reach as high as $\sim 100\,$Jy\,ms.
Assuming the bursts were evenly distributed in  time, we would expect $\sim 4$ bursts in one VAST-P1 observation and therefore a mean flux density of  $\lesssim 1\,$mJy  (based on the observed maximum fluence) in a 12-min image. However, we detected PSR~J1739$-$2521 with a flux density of $\sim 2\,$mJy in four separate observations, and this suggests  that the bursts are not uniformly distributed in time.
\citet{2017ApJ...840....5C} noticed that some RRATs, including PSR~J1739$-$2521, turned ``on'' and ``off'' regularly, with a timescale of  $\sim 30\,$min in the case of PSR~J1739$-$2521.  Therefore if one of the VAST images occurred during an ``on" period, we would expect $\gg 4$ bursts and potentially a flux density consistent with what we measured (depending on the nature of the clustering).
Future observations of RRATs with surveys like VAST can help determine the nature of such  ``on'' and ``off'' behaviour for other sources.

\subsection{Low Mass X-ray Binaries}
\begin{figure}
    \centering
    \includegraphics[width=0.8\columnwidth]{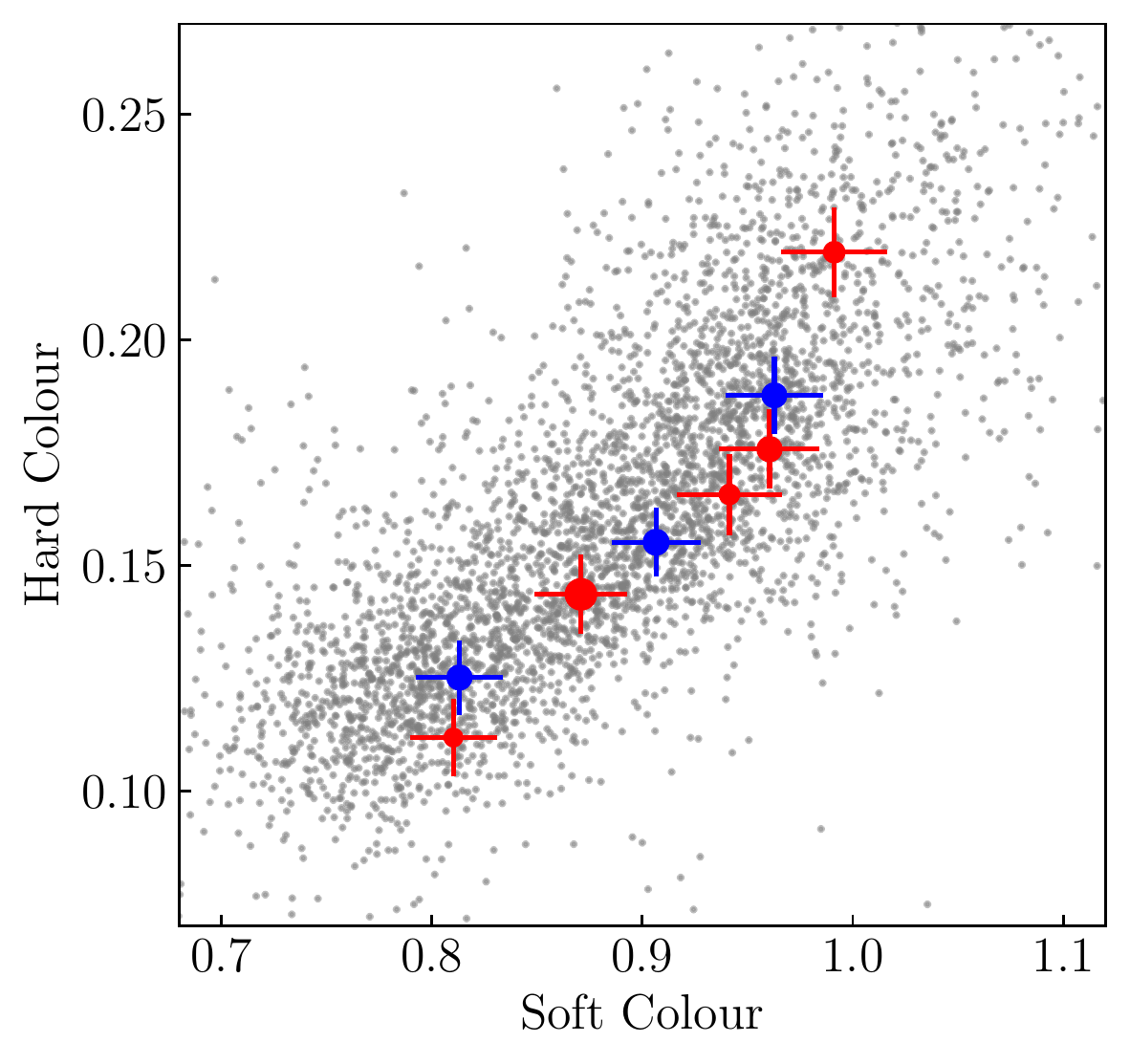}
    \caption{MAXI X-ray colour-colour diagram for 4U 1758$-$25. The ``hard colour'' is the count rate ratio for the energy channel 10--20\,keV to the 4--10\,keV channel; the ``soft colour'' is the ratio for the 4--10\,keV channel to the 2--4\,keV energy channel. Circles show the colours for the closest MAXI observation, where blue circles show the radio detections and red circles show the radio non-detections. The size of circles infers either the peak flux density measurements (for detections) or $3\sigma$ upper-limits of the forced-fitted flux density (for non-detections).}
    \label{fig:GX5_1}
\end{figure}

\textbf{VAST J180108.1$-$250442} is identified as 4U 1758$-$25 (1.7\,arcsec offset), a Z-type LMXB \citep[e.g.,][]{1992ApJ...385..314T}. All Z-type LMXBs have been detected at radio wavelengths and show rapid variability \citep[e.g.,][]{2008MNRAS.390..447T}.
For example, \citet{1992ApJ...385..314T} observed this source with the Australia Telescope Compact Array (ATCA) for four consecutive hours, and they found that the radio emission at 4.9\,GHz could be as high as $\sim$5.8\,mJy but could also be as low as $\sim$1.7\,mJy.
\citet{1988Natur.336..146P} found that the radio emission varied as a function of the position in the X-ray colour-colour diagram, which was associated with changes in the mass accretion rate. 
For example, the radio flux of 4U 1758$-$25 was found to be lowest when the source was on horizontal branch, and highest on the normal branch \citep{1992ApJ...385..314T}.
We searched for X-ray data for this source from an all sky monitoring program, the Monitor of All-sky X-ray Image \citep[MAXI;][]{2009PASJ...61..999M}\footnote{ \url{http://maxi.riken.jp/pubdata/v7l/J1801-250/index.html}}. 
In Figure~\ref{fig:GX5_1}, we show the X-ray colour-colour diagram for 4U 1758$-$25 based on the data from MAXI. The source was likely on the ``normal branch" when detected in VAST, which is associated with radio detections in \citet{1992ApJ...385..314T} and is therefore likely consistent with what we see.  Future observations of a larger sample of LMXBs will help elucidate radio--X-ray state correlations of other sources.

\subsection{Galactic Centre Radio Transients}
Galactic Centre Radio Transients (GCRTs) are a group of unclassified radio sources located towards the Galactic centre. There are three GCRTs detected so far: GCRT~J1746--2757 \citep{2002AJ....123.1497H}, GCRT~J1745--3009 \citep{2005Natur.434...50H} and GCRT~J1742--3001 \citep{2009ApJ...696..280H}. We identified one highly variable source likely to be a new GCRT, which was published separately by \citet{2021ApJ...920...45W}. We summarize the results from that work here.  

\textbf{VAST J173608.2$-$321634} is not coincident with any source at other wavelengths within the surveys we searched. The flux density of the source varied more than a factor of $\sim$10 over timescales of months. The source was also detected to be circularly polarised (up to 41 per cent).  Pulsars are commonly variable in the radio and show circular polarisation. We therefore searched for pulsations with the Parkes telescope (Murriyang) but found nothing. Later, we used the MeerKAT telescope to perform simultaneous imaging and pulsar searching observations. We did not find any pulsations but measured the flux density declining exponentially with a timescale of a day. We calculated the spectral index within the bandpass to be $\sim -3$.
Both this source and GCRTs are highly polarised and with steep spectra.  Overall we could not classify this source as a star, pulsar or other source type, and so its most likely classification is as a GCRT.

We also performed a targeted search for radio emission from the other known GCRTs. 
The estimated flux density at 887.5\,MHz of the burst can reach $\sim$10\,mJy in one 12-min VAST observation (assuming the burst with a peak flux density of $\sim$1\,Jy at 330\,MHz with a $\sim$10\,min duration and a spectral index of $\sim -4$), which can be easily detected ($\sim 20\sigma$) in the VAST-P1.
However, there is no detection from any of these three GCRTs (GCRT~J1746--2757, GCRT~J1745--3009, and GCRT~J1742--3001) with a 5$\sigma$ upper limit of $\sim$5\,mJy (see details in Table~\ref{tab:gcrt}). There are two possible interpretations: (1) the GCRTs were ``off'' in all observations; (2) the spectral index was steeper than that we used to estimate expected flux densities, for example, \citet{2007ApJ...660L.121H} detected a steep-spectrum burst from GCRT~J1745--3009 with a spectral index of $-13.5$ (this would give us an estimated flux density at 887.5\,MHz of $\sim$1\,$\mu$Jy).
Regardless, further VAST observations once the full survey commences will be useful to investigate these sources. Future observations may give us insight into the behaviour of GCRTs at higher frequencies and perhaps find new GCRTs towards the Galactic Centre.

\begin{table*}
    \centering
    \caption{Details of ATCA observations. For each observation, we list its starting date, central frequency, integration time, project code,  and antenna configuration.}
    \begin{tabular}{cccccc}
\hline
Source & Start Date & Central Frequency (MHz) & Time (min) & Project Code & Configuration \\
\hline
VAST~J171631.9$-$303900 & 2021-12-23 & 2100 & 725 & C3363 & 1.5A \\
\hline
VAST~J172841.2$-$334548 & 2021-04-25 & 2100 & 160 & C3431 & 6D \\
 & 2021-06-04 & 2100 & 240 & C3431 & 6B \\
 & 2021-09-11 & 2100 & 300 & C3431 & 6A \\
 & 2021-12-24 & 2100 & 820 & C3363 & 1.5A \\
\hline
VAST~J174655.7$-$245503 & 2021-12-25 & 2100 & 860 & C3363 & 1.5A \\
\hline
VAST~J174917.3$-$204841 & 2020-11-15 & 2100 & 530 & C3363 & 6B \\
 & 2021-12-26 & 2100 & 780 & C3363 & 1.5A \\
 \hline
VAST~J180007.2$-$261251 & 2020-11-14 & 2100 & 510 & C3363 & 6B \\
\hline
\end{tabular}

    \label{tab:ATCA}
\end{table*}

\subsection{Other sources}\label{sec:dis.other}
The remaining four highly variable sources and one highly polarised source were not associated with a previously identified pulsar or star. 
In order to get better positions, we observed these five candidates with the ATCA for 6--10\,hours at 2.1\,GHz. The observation details are listed in Table~\ref{tab:ATCA}.
The typical positional uncertainty for the VAST survey is $\sim 1\,$arcsec \citep[see][for example]{2021PASA...38...54M}, while for our ATCA follow up observations is $\sim0.3\,$arcsec (assuming the configuration with the lowest resolution we used, 1.5A). 
In the ATCA follow-up observations, we only detected three out of five candidates (see the discussion below). 
As ATCA has a better positional accuracy, we will use ATCA positions to find any multi-wavelength counterparts for these three candidates in the following discussion.
In Table~\ref{tab:starmatch}, we summarise the possible optical (\textit{Gaia}) and infrared (VVV) counterparts for each candidates, and calculated the false association rate (FAR) for each potential counterpart to quantify the possibility of spurious associations. The \textit{Gaia} and VVV source in the same row are likely the same source, as the separation between \textit{Gaia} source and VVV source is small ($\lesssim0.6\,$arcsec) and the positions of nearby sources also match.
\correct{Assuming Poisson distribution for sources, }the FAR for an object with an offset $r$ and a magnitude $m$ is given by ${\rm FAR} = 1 - \exp{\left(-\pi r^2 \sigma_{\le m}\right)}$, where $\sigma_{\le m}$ is the average surface density of sources brighter than $m$.
In order to get a robust counterpart, we will only discuss the source with an FAR for VVV $<5$ per cent in detail in the following section. FAR relies on the catalogue completeness, but the completeness for catalogues is lower for crowded regions than other region. The average star density in \textit{Gaia} near our candidates is $\sim8\times10^5$\,deg$^{-2}$. \citet{2021A&A...649A...5F} reported that the completeness of \textit{Gaia} EDR3 catalogue for crowded (star density ranges from $5\times10^5$ to $2\times10^6$ deg$^{-2}$) regions is $\sim20$ per cent lower than the uncrowded (star density less than $5\times10^5$ deg$^{-2}$) ones. To the magnitude limit for our candidates ($\sim$14\,mag in $K_s$-band), the completeness for VVV catalogue is higher than \textit{Gaia} \citep[see e.g.,][]{2019A&A...629A...1S}, though the completeness towards the Galactic Bulge is still lower than that for other uncrowded regions. A low completeness means an underestimate of $\sigma_{\le m}$, which then leads to an underestimate of the FAR.

\begin{table*}
  \centering
  \caption{Highly variable and circular polarised sources with possible \textit{Gaia} (optical) and VVV (infrared) counterparts in our search. ``Best Radio Position'' gives the best radio coordinate of the candidate. For each \textit{Gaia} and VVV counterpart, we list its name (or ID), offset between our best radio position and the \textit{Gaia} or VVV counterpart, and false association rate (FAR) for this  counterpart. In following discussion, we considered the counterpart with an FAR for VVV $<$5 per cent as a true match.
  }
  \label{tab:starmatch}
    \begin{tabular}{ccccccccc}
    \hline
    Source & Best Radio Position & \multicolumn{3}{c}{\textit{Gaia}} & & \multicolumn{3}{c}{VVV} \\
    \cline{3-5}\cline{7-9}
     & & Name & Offset ($''$) & FAR & & Name & Offset ($''$) & FAR \\
    \hline
    VAST~J171631.9$-$303900 & 17:16:31.96 $-$30:39:01.38 & 5980701316456690944 & 0.43 & 0.040 & & ... & ... & ... \\
     & & 5980701312115439616 & 1.48 & 0.023 & & VVV~J171631.92$-$303859.91 & 1.53 & 0.020 \\
    \hline
    VAST~J172841.2$-$334548 & 17:28:41.2 $-$33:45:48 & 5975990832499505152 & 1.79 & 0.009 & & VVV~J172841.33$-$334549.71 & 1.70 & 0.075 \\
    \hline
    VAST~J174655.7$-$245503 & 17:46:55.80 $-$24:55:03.88 & 4068030058829222400 & 0.59 & 0.037 & & VVV~J174655.83$-$245504.24 & 0.61 & 0.021 \\
     & & 4068030024469484160 & 0.93 & 0.057 & & VVV~J174655.73$-$245503.66 & 0.88 & 0.039 \\
    \hline
    VAST~J174917.3$-$204841 & 17:49:17.3 $-$20:48:41 & 4118772073824693504 & 1.90 & 0.040 & & VVV~J174917.20$-$204841.36 & 1.80 & 0.055 \\
    \hline
    VAST~J180007.2$-$261251 & 18:00:07.23 $-$26:12:53.22 & 4064068277987477632 & 0.45 & 0.002 & & VVV~J180007.24$-$261252.67 & 0.58 & 0.000 \\
    \hline
    \end{tabular}
\end{table*}

\begin{figure*}
    \centering
    \includegraphics[width=0.99\textwidth]{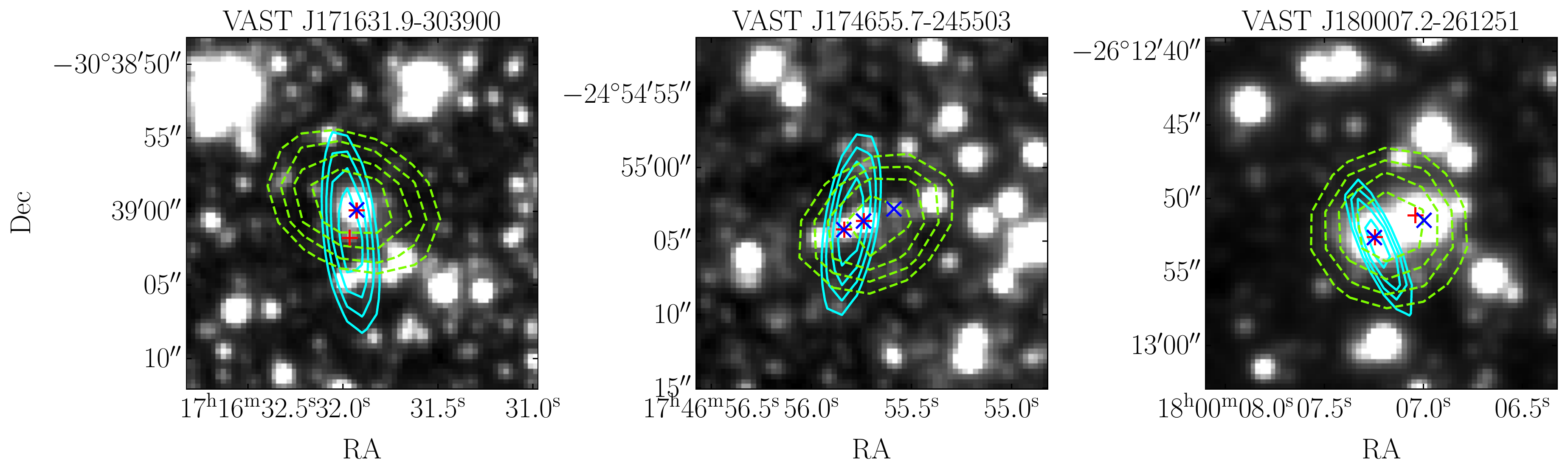}
    \caption{ASKAP (green dashed) and ATCA (cyan solid) contours at 60, 70, 80, 90 per cent of the peak Stokes I flux density overlaid on VVV $J$-band image for three of our candidates detected in ATCA follow-up observations. We also plotted VVV and \textit{Gaia} sources nearby in blue $\times$ and red $+$, respectively.}
    \label{fig:ATCA_VVV}
\end{figure*}

\textbf{VAST J171631.9$-$303900} was detected in our ATCA follow-up observations. We observed this source for 10\,hrs and detected a source at RA = 17:16:31.96$\pm$0.01, Dec = $-$30:39:01.38$\pm$0.61 with a peak flux density of 0.44$\pm$0.04\,mJy at 2.1\,GHz (see Figure~\ref{fig:ATCA_VVV}). 
This source may be coincident with the optical counterpart, \textit{Gaia} 5980701312115439616 with a magnitude $G=17.12$ and a colour $G-G_{\rm RP} = 1.14$, and the infrared counterpart, VVV~J171631.92$-$303859.91 with a magnitude $K_s=12.97$ and a colour $J-K_s=1.04$. 
We can use a distance-independent colour $G - G_{\rm RP}$ \citep[see e.g.,][]{2018A&A...616A...8A} and typical intrinsic colour and magnitude for dwarfs stellar object \citep{2013ApJS..208....9P}\footnote{\url{https://www.pas.rochester.edu/~emamajek/EEM_dwarf_UBVIJHK_colors_Teff.txt}} to infer the source properties. The \textit{Gaia} colour implies an effective temperature of 3500\,K or so (i.e., spectral type of M or so). 
If this object is a dwarf star, the absolute magnitude in $G$-band would be 11\,mag or so, which implies the distance to the source is about 160\,pc (consistent with that derived from the VVV magnitude). The corresponding radio luminosity is $\sim10^{17}$\,erg\,Hz$^{-1}$\,s$^{-1}$, which is consistent with that for typical dwarf stars \citep[e.g.,][]{1995A&AS..109..177W}.
The parallax for such a distance ($\sim$5\,mas) would be easily detected by \textit{Gaia}, but no parallax detected with a $5\sigma$ limit of $\sim 1\,$mas (for sources with a magnitude $G\sim17$). However, the parallax error can reach $\sim$1\,mas for the worst case. 
Based on the VVV colour and magnitude, this source could also be an M-type giant star at a distance of $\sim$2\,kpc, if we using the extinction coefficients in \citet{2013MNRAS.430.2188Y} and assuming an average extinction in the visual band of $A_V/d \approx 1.8\,{\rm mag}\,{\rm kpc}^{-1}$ \citep{1992dge..book.....W}.
Further optical or infrared follow-up could be helpful in determining the precise distance. 
We also identified another possible counterpart in \textit{Gaia} DR3 but with a higher FAR (see Table~\ref{tab:starmatch}) and we will not discuss that in details.

We note that our ATCA follow-up observation had a larger uncertainty in the declination direction. A more precise radio observation would help us to get a better astrometry.
Deeper optical and/or infrared observations are also needed to determine which (if either) infrared source is the counterpart of the radio source and help elucidate the nature of the source.

\begin{figure}
    \centering
    \includegraphics[width=0.95\columnwidth]{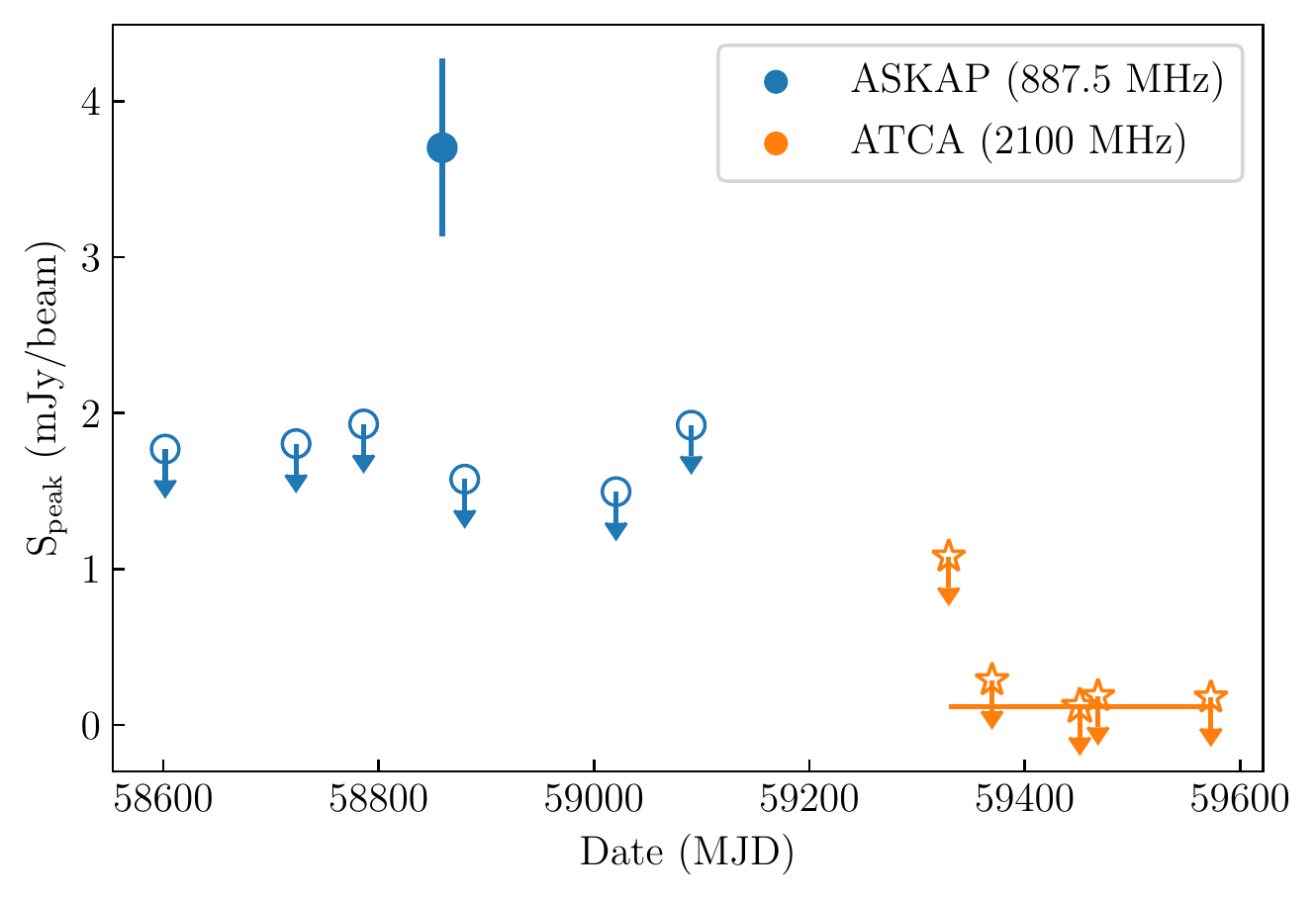}
    \caption{Full radio lightcurve for VAST J172841.2$-$334548. Blue circles show observations with ASKAP while orange stars show observations with ATCA.}
    \label{fig:J1728_lc}
\end{figure}

\textbf{VAST J172841.2$-$334548} was not detected (with a $3\sigma$ upper limit of $<150\,\mu$Jy) in our ATCA follow-up observations for this candidate. We combined all four ATCA follow-up observations and derived a $3\sigma$ upper limit of the quiescent radio emission of $<90\,\mu$Jy (see Figure~\ref{fig:J1728_lc}).
Though the FAR for the possible \textit{Gaia} match is less than 5 per cent, we do not think this object is a reliable counterpart as the source density for \textit{Gaia} in this region is $\sim$20 times less than that for all other candidates, but the source density for VVV is consistent.
There is a probability that there is no detectable optical or infrared counterpart with \textit{Gaia} and/or VVV sensitivity.
This radio source was observed to be $\sim$100 per cent circularly polarised. Sources with circular polarisation but without an obvious infrared or optical counterpart could be cool brown dwarfs \citep[e.g., BDR~J1750+3809, ][]{2020ApJ...903L..33V}, pulsars \citep{2008ApJ...687..262K} or GCRTs \citep{2010ApJ...712L...5R}. 
If the source is a brown dwarf with a similar radio spectral luminosity ($\sim5\times10^{15}\,$erg\,s$^{-1}$\,Hz$^{-1}$, \citealt{2020ApJ...903L..33V}) as BDR~J1750+3809, the source would be at a distance of $\sim33$\,pc. A deeper near-infrared observation would be helpful to find the counterpart and constrain its spectral type.
As noted in \citet{2018MNRAS.474.4629J}, very few ($\lesssim$0.2 per cent) pulsars have fractional circular polarisation $>50$ per cent. If the radio emission came from a pulsar, it could be a pulsar with unusual polarimetric properties.
With a high variability, high fractional circular polarisation and no obvious optical/infrared counterpart, this source could be another GCRT.
A steep radio spectrum is another characteristic of GCRTs. However, we cannot measure a reliable spectral index within the bandpass for the VAST detection due to the low detection significance. Assuming the source was radio bright with the same flux density at low frequency (887.5\,MHz) when we monitored the source with ATCA, the spectral index of the source would be as high as $\sim -4$. However, the spectral index estimation is likely to be wrong as the flux density is highly variable. 
This source was only detected once in our survey. Higher resolution and sensitivity radio monitoring observations would be beneficial in determining the emission cadence, spectral properties and other wavelength counterparts. The same strategy as VAST~J173608.2$-$321634 can be applied to this source. For monitoring observations, simultaneous pulsar searching observations can help us confirm or rule out the pulsar origin.

\textbf{VAST J174655.7$-$245503} was detected in our ATCA follow-up observations at  RA = 17:46:55.80$\pm$0.01, Dec = $-$24:55:03.88$\pm$0.32 with a peak flux density of 0.70$\pm$0.04\,mJy at 2.1\,GHz (see Figure~\ref{fig:ATCA_VVV}).
This source appears coincident with the infrared source VVV J174655.83$-$245504.24 with a colour $J-K_s = 1.70$ and a magnitude $K_s = 14.23$ and the optical source \textit{Gaia} 4068030058829222400 with a magnitude $G = 20.62$ (not detected in $G_{\rm RP}$). The other object (\textit{Gaia} 4068030024469484160 and VVV J174655.73--245503.66) nearby (see Table~\ref{tab:starmatch} and middle panel in Figure~\ref{fig:ATCA_VVV}) could also be the counterpart. We did the same analysis as that for VAST~J171631.9$-$303900. The VVV colours suggest both of them are likely to be a red giant branch stars at a distance of $1-2\,$kpc with a radio luminosity of $\sim10^{18}$\,erg\,Hz$^{-1}$\,s$^{-1}$.
A higher resolution radio observation is needed to determine which object (if either) is the counterpart.

\textbf{VAST J174917.3$-$204841} was not detected (with a $3\sigma$ upper limit of $<100\,\mu$Jy) in our ATCA follow-up observations for this candidate.
As is shown in Table~\ref{tab:starmatch}, the FAR for possible nearby object is greater than 5 per cent and therefore we do not regard this object as a reliable counterpart.
Like VAST~J172841.2$-$334548, this source was only detected once in our survey. Higher resolution radio monitoring observations are needed for determining the source nature.

\textbf{VAST J180007.2$-$261251} was detected in our ATCA observation at RA = 18:00:07.23$\pm$0.01, Dec = $-$26:12:53.22$\pm$0.17 with a peak flux of 0.94$\pm$0.03\,mJy.
Based on the ATCA position, we identified the infrared source VVV J180007.24$-$261252.67  with a colour $J-K_s=1.14$ and a magnitude $K_s=10.68$ as a likely counterpart. 
This source was also identified as \textit{Gaia} 4064068277987477632 with a colour $G-G_{\rm RP}$ = 1.74 and a magnitude $G = 17.83$. 
The source was also identified as a large amplitude variable in \textit{Gaia} DR2 with an amplitude of variation of $\sim$0.3\,mag in $G$-band \citep{2021A&A...648A..44M}.
The \textit{Gaia} colour suggests the source is with an effective temperature of $\sim$2500\,K (i.e., spectral type of M or L). If this object is a dwarf star, the absolute magnitude in $G$-band would be approximately 16\,mag, which implies the distance to the source is only about 30\,pc. The parallax for such a distance ($\sim$30\,mas) would be easily detected by \textit{Gaia}, but no parallax detected with a $5\sigma$ limit of $\sim 5\,$mas. However, the goodness of fitness is poor and excess noise is high for this \textit{Gaia} measurement. The infrared colour ($J-K_s = 1.14$) suggests an extinction in $V$-band of $A_V \sim 1.0$\,mag (assuming a spectral type of late M). This extinction implies that the distance to the source should be around 600\,pc. At such a distance, the source should be $\sim$6\,mag brighter than a dwarf star with the same temperature, which suggests that this object might be a giant star with a radio luminosity of $\sim 10^{18}\,{\rm erg}\,{\rm Hz}^{-1}\,{\rm s}^{-1}$.
In order to get a more reliable distance estimation and therefore work out the source nature, further optical or near infrared follow-up observations are needed. 

In summary, we have searched for possible optical or infrared sources that are coincident with our radio sources. We identified three candidates as possible stellar objects. Two of them are likely to be giant stars and the remaining one might be a dwarf star. Giant stars that are radio loud are usually found in binary systems, such as RS Canum Venaticorum systems \citep[e.g.,][]{2021A&A...654A..21T} and Symbiotic stars \citep[e.g.,][]{1984ApJ...284..202S}. Radio emission from dwarf stars can arise from plasma emission or electron cyclotron maser emission (see \citealt{1985ARA&A..23..169D} for a detailed review). However, with limited optical and infrared data, we can only discuss properties of the object semi-quantitatively based on its colours, magnitudes and distance estimations. A detailed spectroscopy infrared observation can be used to confirm the binary nature and determine the spectral type. 
The remaining two need further investigations such as deep optical and/or infrared observations and further radio monitoring observations, as the possible nearby optical and infrared matches are with a high FAR. 
Our lower limit on the flux ratio of near-IR ($J$-band) to radio ($F_{\rm radio}/F_J$) is $\sim$1, which is more extreme than almost all types of stars (see Figure~7 in \citealt{2022ApJ...930...38W}).  Extinction could moderate this conclusion somewhat, but is unlikely to be a major factor for most main sequence/dwarf stars (see the discussion above). This suggests that if these sources are stellar they probe the extreme end of the radio luminosity distribution which could be probed by deeper near-IR observations, or that they are another source class entirely.
Deeper optical and infrared observations is useful for searching faint counterparts such as ultracool dwarfs. Higher resolution radio monitoring observations would help us to get a better astrometry, know more about the radio emission behaviors and therefore work out the origin of the transient radio emission.

\subsection{Variability rates analysis}\label{sec:rate_discuss}

We found eight highly variable or transient sources out of 29\,410 unique sources within the Galactic Centre region of VAST-P1 with a sky area of 265\,deg$^2$. 
The real number of variable or transient sources could be higher as we adopted strict variability thresholds and excluded sources with siblings or neighbours within 30\,arcsec ($\sim$3 per cent survey area in our search).
We note that the search strategy used in this paper would overlook sources in complex regions such as pulsars in supernova remnants. Different methods such as targeted search or image subtraction will work better for those cases.
Only a small percentage (0.03 per cent) of sources above $\sim$2\,mJy are variable on a timescale of a few months. This is slightly higher than that found in other regions in VAST-P1 \citep{2021PASA...38...54M} but overall they are both consistent with previous searches \citep{2016ApJ...818..105M}.

\begin{table*}
  \centering
  \begin{threeparttable}
  \caption{Properties of GCRTs and GCRT-like candidates. $\Theta$ gives the angular seperation between the source and the GC. We also list the flux density range, variability timescale and polarisation (``C'' stands for circular and ``L'' for linear) information for each source. $\alpha$ gives the radio spectral index for each source. nE gives the number of  epochs that cover the source location. $\sigma_{\rm VAST}$ gives the typical rms for the region near the source in the VAST survey.}
  \label{tab:gcrt}
    \begin{tabular}{ccccccccl}
    \hline
    Source & $\Theta$ & Flux Density & Variability Timescale & Polarisation & $\alpha$ & nE & $\sigma_{\rm VAST}$ & Ref.\\
     & (deg) & (mJy beam$^{-1}$) & & & & & (mJy beam$^{-1}$) & \\
    \hline
    GCRT~J1746--2757 & $\sim$1.0 & $<$20 -- $\sim$200 (330\,MHz) & weeks to months & ... & ... & 10 & 1.2  & 1\\
    GCRT~J1745--3009 & $\sim$1.3 & 15 -- $\sim$2\,000 (330\,MHz) & minutes / repeating & $<$15\% -- $\sim$100\%  C & $-13$ to $-4$ & 7 & 1.0  & 2,3,4,7\\
    GCRT~J1742--3001 & $\sim$1.0 & $<$5 -- $\sim$100 (235\,MHz) & weeks to months & ... & $<-2$ &  9 & 1.3  & 5\\
    C1709--3918 & $\sim$12.7 & 3.2 (618\,MHz) & years to decades$^{\rm a}$ & 14\%  C & $-3.18$ & 1 & 0.8  & 6 \\
    C1748--2827 & $\sim$0.7 & 0.6 (1.26\,GHz) & years to decades$^{\rm a}$ & 14\%  C & $-2.85$ & 10 & 2.5  & 6 \\
    VAST~J173608.2--321634 & $\sim$4.0 & \makecell{$<$1.0 -- 50 (888\,MHz) \\ $<$0.1 -- 5.6 (1.3\,GHz)} & days to weeks & \makecell{30\%  C (888\,MHz) \\ 80\% L (1.3\,GHz)} & $-5.5$ to $-2.6$ & 9 & 0.5 & 8\\
    VAST~J172841.2--334548 & $\sim$5.0 & $<$1.0 -- 3.5 (888\,MHz) & $<$months & $\sim$100\%  C & ... & 7 & 0.6  & this work\\
    VAST~J174917.3--204841 & $\sim$8.0 & $<$1.0 -- 3.1 (888\,MHz) & $<$months & $<$35\%  C & ... & 9 & 0.3 & this work \\
    \hline
\end{tabular}
    \begin{tablenotes}
    \item[a] \citet{2021MNRAS.507.3888H} reported no evidence of significant variable behavior on any timescale and put a constraint on variability of $\lesssim$20\% on year-to-decade timescale.
    \item \textbf{Reference codes:} $^{1}$\citealt{2002AJ....123.1497H}, $^{2}$\citealt{2005Natur.434...50H}, 
    $^{3}$\citealt{2006ApJ...639..348H}, 
    $^{4}$\citealt{2007ApJ...660L.121H}, 
    $^{5}$\citealt{2009ApJ...696..280H}, 
    $^{6}$\citealt{2021MNRAS.507.3888H}, 
    $^{7}$\citealt{2010ApJ...712L...5R}, 
    $^{8}$\citealt{2021ApJ...920...45W}
    \end{tablenotes}
  \end{threeparttable}
\end{table*}


GCRTs are steep spectra polarised radio transients towards the Galactic Centre with no clear other wavelengths counterpart \citep{2005Natur.434...50H, 2008ApJ...687..262K, 2010ApJ...712L...5R}.
We find one source, VAST~J173608.2$-$321634 \citep{2021ApJ...920...45W}, which meets most of the criteria (variable, polarised, steep spectrum, and no counterpart) of the GCRTs.  We find two additional sources that meet some of these criteria, but which also have not been investigated fully.  For instance, VAST~J172841.2$-$334548 is polarised, variable, and has no counterpart at other wavelengths. However, with only one detection we can not determine the lightcurve or spectral index. VAST~J174917.3$-$204841 is variable and with no other wavelengths counterpart. Given the relative low detection significance, we cannot put a strong constraint on spectral indices and fractional circular polarisation estimation. We therefore consider these candidate GCRTs pending further investigations, which should also include deeper optical and infrared observations to determine which (if any) counterparts are plausible. That gives a total number of such unclassified sources of one to three. Other surveys have also found a few GCRT-like sources. For example, \citet{2021MNRAS.507.3888H} found two steep-spectrum polarised source, C1748--2827 and C1709--3918, towards the Galactic Bulge but with no counterpart at other wavelengths. They were thought to be pulsar candidates but no pulsations were detected.
We summarize the properties of the confirmed GCRTs and GCRT-like candidates in Table~\ref{tab:gcrt}.

Calculating transient detection rates for GCRT-like sources can allow us to analyze the source population. 
We used Equation~3 in \citet{2016ApJ...833...11C} to calculate the transient rate $\rho$, the number of detectable transients per unit solid angle per unit time for the highly likely GCRT, VAST~J173608.2$-$321634.
The peak flux density for this source was $12\,$mJy, corresponding to an rms noise threshold of $2.4\,$mJy for a $5\sigma$ detection threshold.
The total volume of our survey $\Omega$ with $\sigma_{\rm rms} < 2.4\,{\rm mJy}$ was $592.3\,{\rm hr}\,{\rm deg}^2$. 
We found one such transient event in the search, which gives us a rate of $\rho(n=1, S_{\rm min}=12\,{\rm mJy}) = 1.7_{-1.6}^{+6.3}\times10^{-3}\,{\rm hr}^{-1}\,{\rm deg}^{-2}$ (the uncertainty was set by 95 per cent confidence intervals). 
The real rate should be slightly higher. Our sparse sampling would miss $\sim 70$ per cent of transient events that can last three weeks (we saw probably persistent emission from VAST~J173608.2--321634 for about three weeks from 2020~Jan~11 to Feb~01). 
The event detection rate for VAST~J173608.2--321634 is comparable to that derived from GCRT~J1745--3009 with a rate of $3.6\times10^{-3}\,{\rm hr}^{-1}\,{\rm deg}^{-2}$ \citep{2016ApJ...832...60P}.
If we assume our source is extragalactic (and hence distributed isotropically on the sky) we could expect to detect $\sim5.6_{-5.3}^{+20.8}$, similar transients in Regions 3 and 4 of VAST-P1. However, no such sources were found in the search \citep[see][]{2021PASA...38...54M}.
The non-detection in this region gives an upper limit of rate of $<2\times10^{-3}\,{\rm hr}^{-1}\,{\rm deg}^{-2}$ at $3\sigma$ confidence. 
Future programs such as the full VAST survey, will help us determine whether GCRT-like sources are distributed isotropically over the sky or mainly distributed towards the GC (see more discussions in $\mathsection$\ref{subsec:future}).


\begin{figure}
    \centering
    \includegraphics[width=0.95\columnwidth]{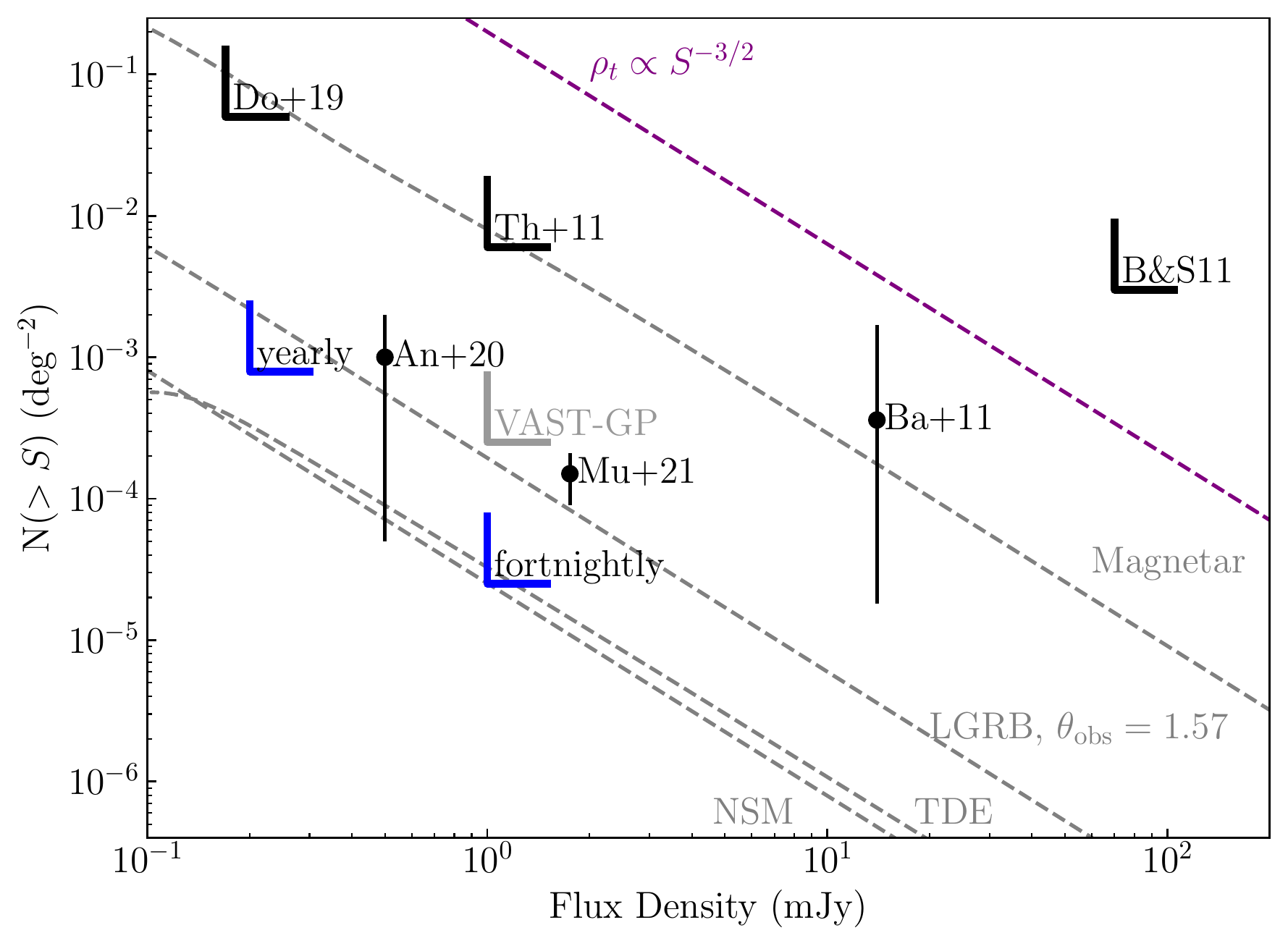}
    \caption{Two-epoch transient source surface density limits adapted from Figure~11 in \citet{2021PASA...38...54M}. Black wedges show upper limits from previous transients search on week-month timescales \citep{2011ApJ...728L..14B, 2011ApJ...742...49T, 2019ApJ...887L..13D}, while black markers show rates from searches with detected transients \citep{2011MNRAS.412..634B, 2020ApJ...903..116A}. 
    Gray wedge denotes upper limit for the originally proposed VAST Galactic Plane field (VAST-GP) from \citet{2013PASA...30....6M}. 
    Blue wedges show the phase space accessible with updated proposed Galactic field in the full survey operations with different search cadences.
    The predicted rates of neutron star mergers (NSM), magnetars, long gamma ray bursts (LGRB) and tidal disruption events (TDE) from \citet{2015ApJ...806..224M} are shown in dashed gray lines. The relation for a Euclidean source population ($\rho_t\propto S^{-3/2}$) is shown with the dashed purple line.}
    \label{fig:tranrate}
\end{figure}

\subsection{Future Prospects} \label{subsec:future}
ASKAP is expected to begin full survey operations in late 2022. The VAST Galactic field region covers a Galactic latitude of $|b| < 6^\circ$ and declination $\delta<0^\circ$, for 42 fields, 1260\,deg$^2$ in total (compared to 5 fields, 265\,deg$^2$ for the VAST-P1 Galactic Centre Region) and the expected cadence for Galactic fields is twice per month for four years.
In Figure~\ref{fig:tranrate}, we show the expected discovery rates for synchrotron transients in new Galactic fields. The phase space we can explore will be hugely improved.
In the coming years, with the VAST full survey, we are able to detect more flare stars, cataclysmic variables, X-ray binaries, pulsars, magnetars and even some previously unknown classes of objects in our Galaxy.
Detections of those unknown classes objects can help us get a more reliable rate estimation and understand their nature. We will be able to catch new outbursts from previously detected GCRTs (at least GCRT~J1745--3009 and VAST~173608.2--321634 are not one-off outbursts) and potentially detect new samples of GCRTs in the full survey. Understanding their distribution would be an essential step of confirming or ruling out their origin. For example, local dwarf star models \citep[e.g.,][]{2010ApJ...712L...5R} would be less likely if they are not distributed isotropically. For the full VAST survey, we are able to observe $\sim$30\,000 hrs (including commensal access from other projects). If we image the data on 12~minutes intervals, we are able to detect $\sim$2\,000 bursts from GCRT-like sources assuming an isotropic distribution, and $\sim$40 bursts assuming a Galactic distribution. The full survey will enable us to put a strong constrain on their distribution and surface density.

\section{Conclusion}\label{sec:sum}

We performed an untargeted search for radio transients and circular polarised sources at 887.5\,MHz using the data from RACS-low and VAST-P1 with ASKAP.

In the transient search, we found eight highly variable sources out of 29\,410 sources in total. These sources included two known pulsars, one low mass X-ray binaries, three stars and two sources yet to be identified. We also discovered one GCRT-like source, whose origin is still unknown. We found an event detection rate for this source of $1.7_{-1.6}^{+6.3}\times10^{-3}\,{\rm hr}^{-1}\,{\rm deg}^{-2}$ which 
is comparable with the rate for GCRT.

In the polarisation search, we found seven highly circular polarised sources out of 4\,278 Stokes V detections in total. Five of them are known pulsars and one of them are the GCRT-like source we discovered in the transient search. We also detected a 100 per cent circular polarised source but have not yet determined its origin. 

With more observations in coming years, we will be able to find more transients towards the Galactic centre. More detections for the transients we discovered in this paper will help us get a better understanding of their origin. Further multi-wavelength follow up observations would also be beneficial in identifying their nature.

\section*{Acknowledgements}
We thank Christian Wolf and Tim Bedding for useful discussions.
TM acknowledges the support of the Australian Research Council through grant DP190100561. DK and AO are supported by NSF grant AST-1816492.
JL and JP are supported by Australian Government Research Training Program Scholarship.
Parts of this research were conducted by the Australian Research Council Centre of Excellence for Gravitational Wave Discovery (OzGrav), project number CE170100004.

This research was supported by the Sydney Informatics Hub (SIH), a core research facility at the University of Sydney.
This work was also supported by software support resources awarded under the Astronomy Data and Computing Services (ADACS) Merit Allocation Program. ADACS is funded from the Astronomy National Collaborative Research Infrastructure Strategy (NCRIS) allocation provided by the Australian Government and managed by Astronomy Australia Limited (AAL).

The Australian Square Kilometre Array Pathfinder is part of the Australia Telescope National Facility which is managed by CSIRO. 
Operation of ASKAP is funded by the Australian Government with support from the National Collaborative Research Infrastructure Strategy. ASKAP uses the resources of the Pawsey Supercomputing Centre. Establishment of ASKAP, the Murchison Radio-astronomy Observatory and the Pawsey Supercomputing Centre are initiatives of the Australian Government, with support from the Government of Western Australia and the Science and Industry Endowment Fund. 
We acknowledge the Wajarri Yamatji as the traditional owners of the Murchison Radio-astronomy Observatory site.
The Australia Telescope Compact Array is part of the Australia Telescope National Facility which is funded by the Australian Government for operation as a National Facility managed by CSIRO.

This research has made use of the SIMBAD database, operated at CDS, Strasbourg, France. 
This research has made use of MAXI data provided by RIKEN, JAXA and the MAXI team.
This research has made use of the VizieR catalogue access tool, CDS, Strasbourg, France. 
This research has made use of NASA's Astrophysics Data System Bibliographic Services.
The acknowledgements were compiled using the Astronomy Acknowledgement Generator. 
This research make use of the following {\sc python} package: {\sc Astropy} \citep{2013A&A...558A..33A, 2018AJ....156..123A}, {\sc matplotlib} \citep{2007CSE.....9...90H}, {\sc NumPy} \citep{2011CSE....13b..22V, 2020Natur.585..357H}, {\sc pandas} \citep{mckinney2010data, jeff_reback_2022_6702671}

\section*{Data Availability}

The ASKAP data used in this paper (RACS-low and VAST-P1) can be accessed through the CSIRO ASKAP Science Data Archive (CASDA\footnote{\url{https://data.csiro.au/dap/public/casda/casdaSearch.zul}}) under project codes AS110 and AS107.
The ATCA data used in this paper can be accessed through the Australia Telescope Online Archive (ATOA\footnote{\url{https://atoa.atnf.csiro.au/query.jsp}}) under project codes C3363.



\bibliographystyle{mnras}
\bibliography{citation} 




\appendix
\section{High False Alarm Rate Counterparts Identification}

In $\mathsection$~\ref{sec:dis.other}, we only discussed counterparts with FAR for VVV less than 5 per cent. We perform the same analysis as what we did for VAST~J171631.9$-$303900 and identify the possible types of nearby objects in Table~\ref{tab:starmatch} with high FAR.

\textbf{VAST J172841.2$-$334548} may be coincident with the infrared source VVV J172841.33$-$334549.71 with a colour $J-K_s=0.98$ and a magnitude $K_s=13.54$ and with the optical source \textit{Gaia} 5975990832499505152  with a colour $G - G_{\rm RP}$ = 1.40 and a magnitude $G = 18.25$. The \textit{Gaia} colour implies an effective temperature of 3000\,K or so (i.e., spectral type of M or so). This object could be a M-type dwarf at a distance of $\sim100\,$pc. Though there is no parallax detection in \textit{Gaia}, the bad goodness of fit and the high significance of excess noise make parallax uncertainty unreliable. Without a more precise astrometry measurement, we cannot rule out this possiblity.

\textbf{VAST J174917.3$-$204841} is coincident with the infrared source VVV J174917.20$-$204841.36 with a colour $(J-K_s)=0.92$ and a magnitude $K_s = 13.34$. This source was also identified as \textit{Gaia} 4118772073824693504 with a colour $G-G_{\rm RP} = 1.03$ and a magnitude $G = 17.00$. The \textit{Gaia} colour implies an effective temperature of 4000\,K or so (i.e., spectral type around K and M). This object could be a M-type dwarf at a distance of $\sim50\,$pc or a giant star at $2-3\,$kpc. Again, a dwarf origin is unlikely as \textit{Gaia} did not detect significant parallax for this object.


\bsp	
\label{lastpage}
\end{document}